\documentclass{myarticle}

\usepackage{inconsolata}

\newcommand{\sM}{\begin{array}{ccccccccc}}
\newcommand{\eM}{\end{array}}

\newcommand{\rM}{{\rm M}}

\renewcommand{\d}{{\,\rm  d}}
\newcommand{\D}{{\,\rm  D}}

\newcommand{\pd}[2]{\displaystyle\frac{\displaystyle\partial #1}{\displaystyle\partial #2}}

\newcommand{\norm}[1]{| #1 |}
\newcommand{\lb}{\left(}
\newcommand{\rb}{\right)}

\newcommand{\sv}{\lb\begin{array}{ccccccccccccccccc}}
\newcommand{\sV}{\begin{bmatrix}}
\newcommand{\eV}{\end{bmatrix}}
\newcommand{\ev}{\end{array}\rb}

\newcommand{\fempty}[1]{{}}

\newcommand{\sty}[1]{\mbox{\boldmath $#1$}}
\newcommand{\styy}[1]{{\mathbb{#1}}}

\newcommand{\fu}{\sty{ u}}

\newcommand{\fx}{\sty{ x}}

\newcommand{\fO}{\sty{ 0}}

\newcommand{\ffC}{\styy{ C}}

\newcommand{\ffR}{\styy{ R}}

\newcommand{\fsigma}{\mbox{\boldmath $\sigma$}}

\newcommand{\feps}{\mbox{\boldmath $\varepsilon $}}

\newcommand{\cC}{{\mathcal{C}}}

\newcommand{\beps}{\ol{\varepsilon}}
\newcommand{\fbeps}{\ol{\feps}}
\newcommand{\fbsigma}{\ol{\fsigma}}

\newcommand{\ol}[1]{\overline{#1}}

\newcommand{\ul}[1]{\underline{#1}}
\newcommand{\ull}[1]{\ul{\ul{#1}}}

\makeatletter
\definecolor{Sblueaa}{cmyk}{1,0.6,0,0}
\definecolor{Sbluea}{cmyk}{1,0.4,0,0}
\definecolor{Sblueb}{cmyk}{0.7,0.2,0,0}
\definecolor{Sbluec}{cmyk}{0.5,0.1,0,0}
\definecolor{Sblued}{cmyk}{0.3,0.05,0,0}
\definecolor{Sbluee}{cmyk}{0.15,0.04,0,0}
\definecolor{Svbluea}{cmyk}{0.9,0.6,0,0}
\definecolor{Svblueb}{cmyk}{0.68,0.4,0,0}
\definecolor{Svbluec}{cmyk}{0.45,0.26,0,0}
\definecolor{Svblued}{cmyk}{0.27,0.12,0,0}
\definecolor{Sblacka}{cmyk}{0.5,0.2,0.2,0.85}
\definecolor{Sblackb}{cmyk}{0.35,0.14,0.14,0.6}
\definecolor{Sblackc}{cmyk}{0.25,0.1,0.1,0.43}
\definecolor{Sblackd}{cmyk}{0.15,0.06,0.06,0.26}
\definecolor{Sblacke}{rgb}{0.827451,0.8509804,0.8627451}
\definecolor{Sred100}{HTML}{EE1C25}
\definecolor{Sorange100}{HTML}{F36F23}
\definecolor{Syellow100}{HTML}{FFDD00}
\definecolor{Spetrol}{HTML}{00AAAD}
\definecolor{Sgreen100}{HTML}{8DC63F}
\definecolor{Spink100}{HTML}{EC008D}
\definecolor{Spurple100}{HTML}{812A91}
\definecolor{Syellow}{cmyk}{0,0.1,1,0}
\definecolor{Sorange}{cmyk}{0,0.7,1,0}
\definecolor{Sred}{cmyk}{0,1,1,0}
\definecolor{Spink}{cmyk}{0,1,0,0}
\definecolor{Spurple}{cmyk}{0.6,1,0,0}
\definecolor{Scyan}{cmyk}{1,0,0.4,0}
\definecolor{Sgreen}{cmyk}{0.5,0,1,0}
\definecolor{Sgreen}{cmyk}{0.5,0,1,0}
\definecolor{uniSgreen}{HTML}{93FF00}
\definecolor{uniSred}{HTML}{FF000B}
\definecolor{uniSpink}{HTML}{FF0098}
\definecolor{uniSorange}{HTML}{FF5D00}
\definecolor{uniScyan}{HTML}{00FBFF}

\colorlet{Sblackf}{Sblacke!70!white}
\colorlet{Sdgreen}{Sgreen!50!black}
\colorlet{Slred}{Sred!40!white}

\usepackage[ruled,lined]{algorithm2e}
\SetAlgoSkip{bigskip}
\LinesNumbered
\g@addto@macro{\@algocf@init}{\SetKwInOut{Input}{Input}}
\g@addto@macro{\@algocf@init}{\SetKwInOut{Output}{Output}}
\g@addto@macro{\@algocf@init}{\SetKwInOut{Requirements}{Requirements\!\!\!\!}}

\definecolor{uniSgray}{RGB}{62, 68, 76}
\colorlet{uniSgray90}{uniSgray!90!white}
\colorlet{uniSgray80}{uniSgray!80!white}
\colorlet{uniSgray70}{uniSgray!70!white}
\colorlet{uniSgray60}{uniSgray!60!white}
\colorlet{uniSgray50}{uniSgray!50!white}
\colorlet{uniSgray40}{uniSgray!40!white}
\colorlet{uniSgray30}{uniSgray!30!white}
\colorlet{uniSgray20}{uniSgray!20!white}
\colorlet{uniSgray10}{uniSgray!10!white}
\definecolor{Scyanlight}{rgb}{ 0.53333,0.87059,0.87451}
\definecolor{uniSblue}{HTML}{004191}
\colorlet{uniSblue80}{uniSblue!80!white}
\colorlet{uniSblue60}{uniSblue!60!white}
\colorlet{uniSblue40}{uniSblue!40!white}
\definecolor{uniSlightblue}{HTML}{00BEFF}
\colorlet{uniSlblue80}{uniSlightblue!80!white}
\colorlet{uniSlblue60}{uniSlightblue!60!white}
\colorlet{uniSlblue40}{uniSlightblue!40!white}
\definecolor{FFgreen}{rgb}{0.635,0.8275,0.1255}
\definecolor{FFdgreen}{rgb}{0.45,0.61,0.09}

\newcommand{\Stilde}{\raise.17ex\hbox{$\scriptstyle\sim$}}

\usepackage{tikz}
\usetikzlibrary{shapes.misc, positioning,arrows,decorations.markings, calc}

\makeatother
\newcommand{\reals}{\mathbb{R}}

\newcommand{\rmdev}{{\mathrm{dev}}}
\newcommand{\sd}{{\mathrm{sd1}}}
\newcommand{\sdd}{{\mathrm{sd2}}}
\newcommand{\eps}{\varepsilon}
\newcommand{\sfT}{\mathsf{T}}

\newcommand{\rmmean}{\mathrm{mean}}

\newcommand{\rmann}{\mathrm{ANN}}
\newcommand{\mse}{\mathrm{MSE}}
\newcommand{\nf}[1]{{\| #1 \|}}

\newcommand{\rmrca}{\mathrm{RCA}}
\newcommand{\rmacc}{\mathrm{ACC}}
\newcommand{\bueps}{{\bar{\underline{\varepsilon}}}}
\newcommand{\busig}{{\bar{\underline{\sigma}}}}
\newcommand{\rn}{{\mathrm{R}N}}

\newcommand{\uT}{\underline{T}}

\newcommand{\ub}{\underline{b}}

\newcommand{\ud}{\underline{d}}

\newcommand{\up}{\underline{p}}

\newcommand{\ux}{\underline{x}}
\newcommand{\uy}{\underline{y}}

\newcommand{\uuA}{\underline{\underline{A}}}

\newcommand{\uuU}{\underline{\underline{U}}}

\newcommand{\uuW}{\underline{\underline{W}}}

\newcommand{\uxi}{\underline{\xi}}

\newcommand{\usigma}{\underline{\sigma}}

\newcommand{\uvarepsilon}{\underline{\varepsilon}}

\newcommand{\usig}{\usigma}
\newcommand{\ueps}{\uvarepsilon}

\newcommand{\rmD}{\mathrm{D}}

\newcommand{\rmF}{\mathrm{F}}

\newcommand{\rmM}{\mathrm{M}}

\newcommand{\rmR}{\mathrm{R}}

\newcommand{\rmT}{\mathrm{T}}

\newcommand{\rmV}{\mathrm{V}}

\newcommand{\bri}{{(i)}}

\makeatletter
\newcommand{\setword}[2]{%
  \phantomsection
  #1\def\@currentlabel{\unexpanded{#1}}\label{#2}%
}
\makeatother
\begin{document}

\title{On-the-fly adaptivity for nonlinear twoscale simulations using artificial neural networks and reduced order modeling}
\author{
Felix Fritzen$^{1}$\footnote{%
Corresponding author
, email: \email{fritzen@mechbau.uni-stuttgart.de}%
, ORCID: \url{https://orcid.org/0000-0003-4926-0068}
}
, Mauricio Fernández$^{1}$%
\footnote{%
email: \email{mauricio.fernandez@mechbau.uni-stuttgart.de}%
, ORCID: \url{http://orcid.org/0000-0003-1840-1243}
}
, Fredrik Larsson$^{2}$
}
\affil{%
$^{1}$Institute of Applied Mechanics (IAM), University of Stuttgart, Pfaffenwalring 7, 70569 Stuttgart, Germany\\
$^{2}$Material and Computational Mechanics, Department of Industrial and Materials Science, Chalmers University of Technology, G\"{o}teborg, Sweden
}
\date{Accepted for publication in Frontiers in Materials (April 5th, 2019)}
\maketitle

\begin{abstract}
A multi-fidelity surrogate model for highly nonlinear multiscale problems is proposed. It is based on the introduction of two different surrogate models and an adaptive on-the-fly switching. The two concurrent surrogates are built incrementally starting from a moderate set of evaluations of the full order model. Therefore, a reduced order model (ROM) is generated. Using a hybrid ROM-preconditioned FE solver, additional effective stress-strain data is simulated while the number of samples is kept to a moderate level by using a dedicated and physics-guided sampling technique. Machine learning (ML) is subsequently used to build the second surrogate by means of artificial neural networks (ANN). Different ANN architectures are explored and the features used as inputs of the ANN are fine tuned in order to improve the overall quality of the ML model. Additional ANN surrogates for the stress errors are generated. Therefore, conservative design guidelines for error surrogates are presented by adapting the loss functions of the ANN training in pure regression or pure classification settings. The error surrogates can be used as quality indicators in order to adaptively select the appropriate -- i.e. efficient yet accurate -- surrogate. Two strategies for the on-the-fly switching are investigated and a practicable and robust algorithm is proposed that eliminates relevant technical difficulties attributed to model switching.  The provided algorithms and ANN design guidelines can easily be adopted for different problem settings and, thereby, they enable generalization of the used machine learning techniques for a wide range of applications. The resulting hybrid surrogate is employed in challenging multilevel FE simulations for a three-phase composite with pseudo-plastic micro-constituents. Numerical examples highlight the performance of the proposed approach.
\end{abstract}

\newpage
\tableofcontents

\newpage
\section{Introduction}

In computer-assisted materials design and in the simulation of complex materials with rich microstructure major challenges remain to be solved despite the outstanding advances made in recent years. For example, the discretization of all microstructural features in a monolithic finite element (FE) simulation is unfeasible due to the various length scales involved that range from micrometers up to the meters. These would lead to a ludicrous complexity of the resulting overall model. By accounting for a separation of length scales, the FE$^2$ ansatz \citep{feyel1999,miehe2002} can lead to some savings over the monolithic approach by replacing the heterogeneous material by microscopic FE problems at the macroscopic integration points, leading to a partial decoupling of microscopic and macroscopic degrees of freedom. Still the number of overall unknowns is prohibitive and calls for massively improved computational efficiency in terms of CPU time, memory savings and information compression. Novel strategies contributing to the vision of a fully connected investigation of materials and aspiring the prediction of process-structure-property relationships across multiple length and time scales are, thus, much sought-after, see, e.g., \cite{Schmitz2016}. Due to the rapid growth of available material and simulation data, data-integrated approaches that exploit information from different sources in order to complement or substitute simulations and experiments are experiencing increased attention, see, e.g., \cite{Kalidindi2015}, \cite{Kalidindi2015c} and \cite{Ramakrishna2018}.

Due to novel improvements in machine learning and computational resources, a zoo of data-driven methods comprising, e.g., kernel methods, principal component analysis and artificial neural networks, have developed immense momentum over the last years. The successful implementation of these techniques in materials research is an active field. For instance, in \cite{Chupakhin2017} artificial neural networks and finite element computations have been combined in order to predict the influence of plasticity on the residual stress field measured by hole drilling. Principal component analysis of $n$-point microstructure statistics have shown excellent performance in order to examine microstructure-property relationships, see, e.g., \cite{Cecen2014}, \cite{Gupta2015} and \cite{Altschuh2017}. In \cite{Belisle2015}, several machine learning methods have been considered in the context of molecular dynamics. \cite{Liu2015a} show how data mining and machine learning are combined in order to efficiently approximate the elastic localization in voxelized microstructures. Another branch of data-driven materials research exploring the use of convolutional neural networks and deep learning in order to deliver accurate structure-property linkages is currently in heavy development, see, e.g., \cite{Cecen2018} and \cite{Yang2018}. 

While data-driven approaches have their appeal, the structure of the underlying physical problem can be accounted for only in parts. For instance, established balanced laws and thermodynamic principles are hard to be incorporated in the aforementioned methods. Reduced order models for the microscopic problem offer an advantageous compromise between physics-informed modeling and computational efficiency. Purely data-driven surrogates lack accuracy (i) if the amount of training data is limited, (ii) if the validity domain is left or (iii) if the error of the surrogate in respect to the reference solution is to be estimated. In these scenarios, reduced order models following physical principles offer, in general, better accuracy and robustness. For example, in \cite{Fritzen2013a} a highly efficient potential based reduced order model has been developed. This ansatz has a natural physical supporting argument, since a reduced basis for the solution field is generated based on snapshot data of FE computations. The approach has been demonstrated to achieve substantial speed-ups and memory savings, see also \cite{Fritzen2014} and \cite{Fritzen2016}. Other developments in this field comprise the NTFA~\citep[e.g.,][]{Michel2003a} and NTFA-TSO~\citep{michel2016b} or hyper-reduced simulations and related schemes \citep{ryckelynck2009b,Soldner2017}. In order to improve the incorporation of surrogates obtained from reduced order models a goal-oriented error estimation or quality indication is required. The quantity of interest (QoI) is the effective stress and its accuracy (up to a prescribed tolerance) is essential for reliability of the overall predictions. In \cite{Lu2018}, for example, neural networks have been successfully trained to approximate the microscopic nonlinear microscopic electric material law of graphene/polymer nanocomposites, but without error control or model adaptivity in macroscopic simulations. A macroscopic goal-oriented approach combining reduced order modeling and machine learning techniques has been demonstrated in \cite{Trehan2017} for two-dimensional oil-water subsurface flow systems and in \cite{Freno2019} for three-dimensional mechanical problems. The approach considers a reduced order model and an a~posteriori correction through machine learning methods. The ansatz shows promising results, but it requires the evaluation of the reduced order model. For twoscale simulations with a macroscopic and a microscopic problem, even the evaluation of a reduced order model for the microscopic problem may not be always viable due to the large number of needed evaluations in the macroscopic problem. It is, therefore, necessary to seek efficient alternatives incorporating a hierarchy of surrogate models of different computational complexity and different accuracy in the QoI. Hereby, one may also take into consideration physics-informed artificial neural networks, as done in \cite{Raissi2018}, in order to obtain the field solution of the balance equations for the microscopic physical problem at hand and then computing the QoI for the macroscopic scale. Such approaches are highly attractive, but not suitable for the objectives of twoscale computations, since the surrogate for the microscopic model is only required to return the QoI for the macroscale and additional calibration of the artificial neural networks for the microscopic solution field would only increase the computational costs without any benefits for the convergence of the macroscopic problem.

The present work aims in mechanical multiscale FE simulations at the adaptive combination of the physics-informed reduced order model (ROM) of \cite{Fritzen2018}, for nonlinear hyperelastic problems (i.e., no history dependency), with artificial neural networks (ANNs), for which feedforward neural networks are considered. Hereby, the ANNs are trained based on FE computations of the full three-dimensional microstructure and material at hand for a set of loading strains (input quantity). The QoI (output quantity) is the effective stress, which the ANNs are trained for. The trained ANNs are then used as a highly efficient constitutive relation surrogate for the nonlinear material at hand. For macroscopic FE computations, the ANN material law surrogate is to be used, if possible, at every integration point for given effective strain. Based on quality indicators, as accuracy or range of validity, the ANN constitutive relation surrogate may be inaccurate or insecure for given strain. The present work, therefore, further considers the error modeling of trained ANNs and discusses guidelines in order to induce conservative properties to the error models, which are calibrated through ANNs in standard regression or classification approaches. Additionally, strategies are proposed for an adaptive ANN-ROM schemes, where the more accurate but expensive ROM is only called at an integration point, if the quality indicator demands it. 

The manuscript is organized as follows: In Section~\ref{sec_romann}, two concurrent surrogate models for the QoI obtained by reduced order modeling and from purely data-driven ANNs are described. The twoscale mechanical problem is introduced and the challenges in the goal-oriented error estimation of derived quantities of interest remaining in nonlinear reduced order modeling are detailed. Then, the data generation for the training of the ANNs is illustrated, followed by the guidelines for the material law and error approximation. At the end of the section, adaptive twoscale simulation strategies including on-the-fly model switching are presented. Section~\ref{sec_nex} offers numerical examples for a three-phase pseudo-plastic material: The ANN is used for the direct surrogation of the QoI. This is adaptively complemented by a more robust and reliable reduced order model based on the concept of quality indicators. Multiscale FE simulations comparing the different multiscale simulation techniques are presented. The manuscript ends with a concluding summary of the results in Section~\ref{sec_disscussion}.

\section{Reduced Order Modeling and Artificial Neural Networks}
\label{sec_romann}
% FF1

\subsection{Twoscale Framework}
\label{sec:Twoscale}
 
\subsubsection{Problem Setting}
The simulation of microstructured solids with a sufficient separation of length scales is investigated. More precisely, a macroscopic domain $\ol{\varOmega}\subset\ffR^3$ with characteristic length~$\overline{L}$ and an attached microstructure with characteristic length~$L_\mu \ll \overline{L}$ are considered. The microstructure is assumed to be ergodic and the existence of a periodic Representative Volume Element (RVE)~$\varOmega$ is assumed. In the following, macroscopic fields are overlined~$\ol{\bullet}$. The twoscale problem consists of the concurrent solution of the macroscopic boundary value problem (BVP)
\begin{align}
 {\rm ( \ol{P})}:& \ {\rm div}\lb \fbsigma(\fbeps)\rb = \fO \quad \text{with} \quad  \fbeps = {\rm sym\;grad} (\ol{\fu}) \quad \text{+ BC}
  \label{eq:macroproblem}
\end{align}
and, for each macroscopic point~$\ol{\fx}\in\ol{\varOmega}$, of the solution of the RVE problem
\begin{align}
 {\rm (P)}:&\ {\rm div}\lb \fsigma(\feps) \rb = \fO \quad \text{with} \quad \feps={\rm sym\; grad}( \fu ) \quad \text{and}\quad 
 \frac{1}{\vert\varOmega\vert} \intop_{\varOmega} \feps \, \d V = \fbeps.
 \label{eq:RVEproblem}
\end{align}
Here $\fu, \ol{\fu}$ denote displacements, $\feps, \fbeps$ are infinitesimal strain tensors and $\fsigma, \fbsigma$ denote the stress fields on the microscopic and macroscopic domain, respectively. The solution of $\rm (P)$ defines the missing constitutive relation for the macroscopic stress~$\fbsigma$ via the volume average
\begin{align}
 \fbsigma &= \frac{1}{\vert\varOmega\vert} \intop_\varOmega \fsigma \, \d V.
 \label{eq:eff:stress}
\end{align}
The two BVPs are strongly coupled since the solution~$\ol{\fu}$ of $\rm (\ol{P})$ defines the boundary condition for $\rm (P)$ via~$\fbeps$, while the solution of $\rm (P)$ implicitly provides the missing constitutive equation via~\eqref{eq:eff:stress}.

\begin{figure}[h]
\centering
\includegraphics[scale=1]{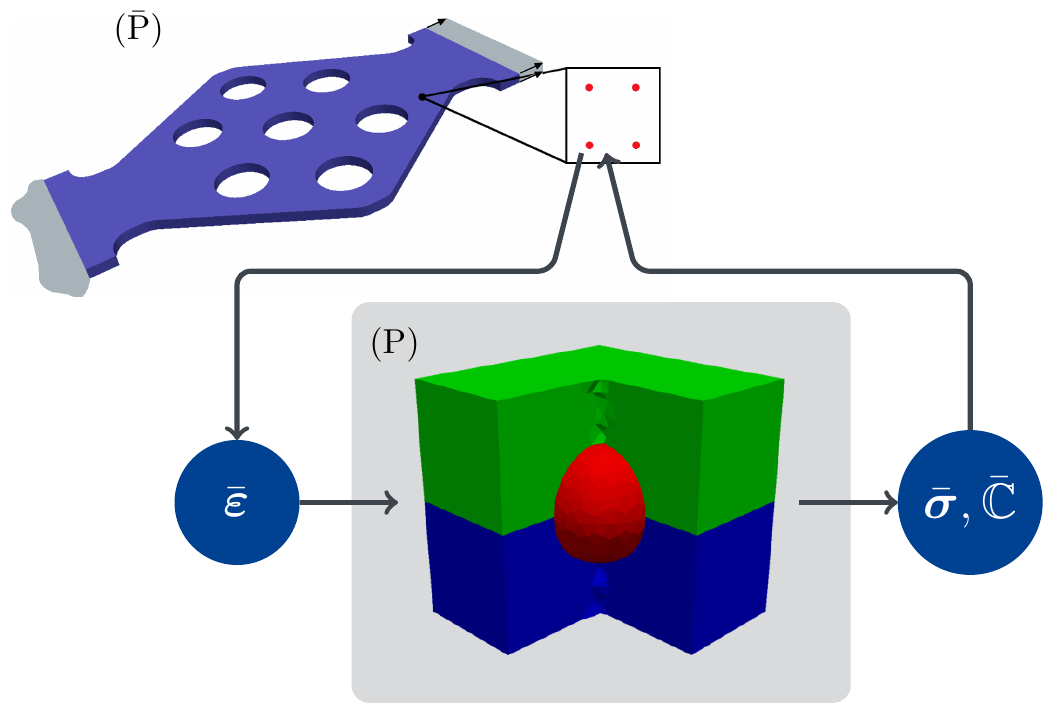}
\caption{Twoscale mechanical problem in the context of FE\textsuperscript{2} for a material with a microstructure composed of three material phases: at every integration point of the macroscopic problem $(\bar{\mathrm{P}})$ the microscopic problem $(\mathrm{P})$ is solved in the FOM through a microscopic FE computation for prescribed $\bar\feps$, the resulting effective stress $\bar\fsigma$ and corresponding gradient $\bar\ffC = \partial \bar\fsigma / \partial \bar\feps$ are then computed and returned to the macroscopic integration point.}
\label{fig_fe2}
\end{figure}

A straight-forward yet computationally costly approach to solving the twoscale problem is given in terms of the FE\textsuperscript{2} method \citep{feyel1999,miehe2002}: Here the microscopic problem is solved at each macroscopic integration point and the effective tangent operator is used in order to allow for Newton-Raphson iterations of the nonlinear macroscopic BVP. In the following, the solution of the microscopic BVP using Finite Elements is considered as the reference  solution, i.e. it denotes the Full Order Model (FOM). In Fig.~\ref{fig_fe2} the macroscopic and microscopic problems are illustrated in the context of FE\textsuperscript{2}.

\subsubsection{Reduced Order Model (ROM)}
\label{sec:ROM}
Given the massive computational demands of the FE\textsuperscript{2} technique and the limited availability of computational resources, the use of nowadays established reduced order models (ROM), in order to replace the costly microscopic BVP evaluations, has become an accepted alternative for dissipative and pseudo-plastic hyperelastic materials~\citep{radermacher2016,Fritzen2018}. The reduced basis of dimension~$N$ obtained from the snapshot Proper Orthogonal Decomposition \citep[POD,][]{sirovich1987} can be expressed in terms of a matrix $\uuU(\ux)$, where each column represents a displacement field. The reduced parameterization of the solution is then given in vector notation via ($i= 1,\dots, N$)
\begin{align}
 \ul{u}(\ul{x}) & = \ull{\beps} \, \ul{x} + \ull{U}(\ul{x})\,\ul{\xi}, &
 \ul{\eps}(\ul{x}) &= \ul{\beps} + \ull{E}(\ul{x})\,\ul{\xi}, &
 \lb\ull{E}(\ul{x})\rb_{\bullet i} &= {\rm vec}\lb{\rm sym\;grad}\lb \lb\ull{U}(\ul{x})\rb_{\bullet i} \rb \rb,
  \label{eq:disp_parametrization}
\end{align}
where $(\uuA)_{\bullet i}$ refers to the $i$th column of the corresponding matrix.
Here, the matrix and vector notation of the effective strain $\fbeps$ are used concurrently for convenience. In the following, attention is limited to pseudo-plastic materials, i.e. to strongly nonlinear hyperelastic solids for which the stress~$\fsigma$ and the stiffness~$\ffC$ are defined as the gradients of a free energy function~$W(\feps)$ according to
\begin{align}
 \usigma \equiv \fsigma & = \pd{W(\feps)}{\feps}, &
 \ull{C} \equiv \ffC & = \pd{^2 W(\feps)}{\feps\, \partial \feps}.
\end{align}
Following \cite{Fritzen2018} the reduced problem is to find the coefficients~$\ul{\xi}\in\ffR^N$ solving
\begin{align}
 \ul{r} (\ul{\beps}, \ul{\xi}) &= \intop_\varOmega \ull{E}^{\sf T} \ul{\sigma} \, \d V \overset{!}{=} \ul{0}. \label{eq:red:res}
\end{align}
While the effective stress is obtained from simple volume averaging of~$\usigma$, the effective tangent stiffness is computed via
\begin{align}
 \ol{\ull{C}} &= \frac{1}{\vert\varOmega\vert} \lb \intop_\varOmega \ull{C} \d V
  - \ull{K}^{\sf T} \, \ull{J}^{-1} \ull{K} \rb &
  \quad \text{with}\quad
  \ull{K} &= \intop_{\varOmega} \ull{C} \, \ull{E} \, \d V, &
  \ull{J} &= \intop_{\varOmega} \ull{E}^{\sf T} \ull{C} \, \ull{E} \, \d V,
  \label{eq:red:jacobianetc}
\end{align}
which follows from straight-forward linearization of \eqref{eq:red:res}. The accuracy of the ROM depends on the quality and amount of the snapshots and of the reduced dimension~$N$. It shall be noted that the Galerkin ROM inherits the properties of classical Finite Elements, i.e. the solution is Galerkin orthogonal and, thus, energy optimal. It follows the basic physical principle of energy minimization. From a theoretical perspective this motivates the robustness and accuracy of the ROM even beyond the considered parameters used during the generation of the snapshot data, i.e., the ROM can be considered to generalize.

% FL
\subsection{Goal-oriented error estimation}
\label{sec:error:theory}

For the microscopic BVP, using the ROM (or any other approximation of the FOM) naturally introduces an error into the solution of the problem, and into the quantity of interest (QoI). In this work the latter is the effective stress. Hence, in order to enable error control for the macroscopic boundary value problem, it is crucial to estimate the error in the QoI, see, e.g., \cite{larssonrunesson2011}. For this purpose, we define the error in displacements on the microscale as ${\ul{e}}(\ux)={\ul{u}}^{\rm F}(\ux)-{\ul{u}}^{{\rm R}N}(\ux)$, where $\ul{u}^{\rm F}$ and $\ul{u}^{{\rm R}N}$ are the solutions to the microscopic problem \eqref{eq:RVEproblem} using the FOM and $N$-dimensional ROM, respectively. In view of the reduced kinematics \eqref{eq:disp_parametrization}, we can parameterize the error in the ROM  as
\begin{align}
    \ul{e}(\ux)={\ull{U}}^{\rm F}(\ux)\uxi_{e},
    \label{eq:error:solution}
\end{align}
where $\ull{U}^{\rm F}$ denotes the (finite element) shape functions pertinent to the FOM and $\ul{\xi}_e$ denotes the nodal values of the fully resolved error. The FOM and $N$-dimensional ROM effective stress functions are referred to for clarity as
\begin{align}
	\busig^\rmF(\bueps) 
	&= \text{effective stress of FOM for given effective strain } \bueps
	\label{eq:sigF:def}
	\ , \\
	\busig^\rn(\bueps)
	&= \text{effective stress of $N$-dimensional ROM for given effective strain } \bueps
	\ ,
\end{align}
while the corresponding error is addressed as
\begin{align}
    \ul{e}_{\ol{\sigma}}=\ol{\ul{\sigma}}^{\rm F}-\ol{\ul{\sigma}}^{{\rm R}N}.
    \label{eq:error:qoi}
\end{align}
Now, consider the corresponding FOM residual equation analogous to \eqref{eq:red:res} and the error in the QoI given in \eqref{eq:error:qoi} in terms of the error in the solution defined in \eqref{eq:error:solution}. Through linearization, we obtain the error equation and the linearization of the macroscopic stress error
\begin{align}
    \ull{J}^{\rm F} \ul{\xi}_e \approx - \ul{r}^{\rm F}, \quad \quad \quad
    \ul{e}_{\ol{\sigma}} \approx \ull{K}^{\rm F}\ul{\xi}_e,
    \label{eq:linearizations}
\end{align}
respectively. Here $\ul{r}^{\rm F}$, $\ull{J}^{\rm F}$ and $\ull{K}^{\rm F}$ define the residual, Jacobian and linearized stress error in \eqref{eq:red:res} and \eqref{eq:red:jacobianetc}, with $\ull{E}$ replaced by $\ull{E}^{\rm F}$ defining the strains of the finite element shape functions. The, nowadays, standard method of goal-oriented error estimation can be carried out by solving the suitably formulated dual (or adjoint) problem \citep[see, e.g.,][]{odenprudhomme2001}
\begin{align}
    \left(\ull{J}^{\rm F}\right)^{\rm T} \ull{\xi}^* = \left(\ull{K}^{\rm F}\right)^{\rm T}.
    \label{eq:dualproblem}
\end{align}
Finally, \eqref{eq:linearizations} and \eqref{eq:dualproblem} can be combined to yield the result 
\begin{align}
    \ul{e}_{\ol{\sigma}} \approx -[\ull{\xi}^*]^{\rm T}\ul{r}^{\rm F}.
    \label{eq:estimator}
\end{align}
We note that the estimator \eqref{eq:estimator} has, in particular, the following properties: (i) It is restricted to estimating the linearized error contribution, (ii) it requires the assembly of the entire FOM residual and Jacobian, and (iii) it requires the solution of the dual problem using the FOM to formally hold. Even if the linearization error is negligible, the high computational cost involved in assembling the full (FOM) Jacobian and residual of the problem makes this technique unalluring for use in conjunction with highly efficient ROM approximations. Possible approximations of \eqref{eq:dualproblem} pertain to hierarchical approximations. One could, for instance, solve the dual problem using an enriched ROM, rather than the FOM. However, designing a robust hierarchical scheme requires means of guaranteeing that the enriched basis is sufficient. In view of the discussion above, we shall henceforth consider alternative methods to estimating (and controlling) the error in macroscopic stress from each microscopic problem.

% MF
\subsection{Artificial Neural Networks (ANNs)}
\label{sec:ANN}
\subsubsection{Generation of data}
\label{sec:data:generation}
\paragraph{Design of input data / loading directions}

The present work is concerned with materials based on state dependent models for, e.g., pseudo-plasticity. For such material models, see, e.g., \citet{kunc2018a}, the transition zone between elastic and plastic domain is found in the vecinity of the origin in strain space. In this transition zone a pronounced nonlinearity and change of slope not only from the elastic to the plastic domain take place, but also depending on the load direction in the plastic domain, followed by a saturation behavior for increasing load amplitudes. This material behavior motivated the Concentric Sampling (CS) approach proposed by \citet{kunc2018a} for pseudo-plastic materials, which is also used in this work. Based on the CS approach, $n_d$ almost uniformly distributed unit vectors / directions $\ud^\bri \in \ffR^6$ ($i=1,\dots,n_d$) are generated. Samples along the generated directions are considered with an exponentially growing step width from the origin. The primal strain dataset~$\hat\rmD_\varepsilon$ is addressed as
\begin{equation}
	\hat\rmD_\varepsilon 
	= \{\bueps \in \reals^6: 
	\bueps = r \ \ud
	, r \in \rmD_r
	, \ud \in \rmD_d
	\} 
	\ , \quad
	\rmD_r
	= \{r_1,\dots\}
	\ , \quad
	\rmD_d = \{\ud_1,\dots\}
	\label{eq:eps:primal}
\end{equation}
with the primal strain norm discretization $\rmD_r$ and set of directions $\rmD_d$. The definition \eqref{eq:eps:primal} corresponds to a tensor decomposition into direction and amplitude. For many materials the volume changes are rather small compared to isochoric deformations. This effect is particularly pronounced for (pseudo-)plastic materials. In order to sample the strain space in a problem specific manner, a rescaling of the strains defined in \eqref{eq:eps:primal} may be convenient. The present work solely rescales the spherical  part (sph) of each primal strain (i.e. the dilatation), while the  deviatoric part (dev) remains unchanged. The actual strain dataset is described by
\begin{equation}
	\rmD_\varepsilon = 
	\left\{
	\bueps \in \reals^6 :
	\bueps 
	= \hat{\uT}(\hat\bueps) 
	= \frac{1}{\hat{r}}\mathrm{sph}(\hat\bueps) + \rmdev(\hat\bueps)
	, \hat\bueps \in \hat\rmD_\varepsilon	
	\right\}
	\ , \quad
	\#(\rmD_\varepsilon) = \#(\rmD_d) \#(\rmD_r)
	\ ,
	\label{eq_dataset_strain}
\end{equation}
where $\hat{r}$ specifies the rescaling of the spherical part. The number of strain samples $\#(\rmD_\varepsilon)$ is given by the product of number of the directions $\#( \rmD_d)$ and the number of amplitudes per direction $\#(\rmD_r)$.

\paragraph{Generation of output data}
\label{sec:gen:output}

For the training of the artificial neural networks (ANNs), training (T), validation (V) and random (Monte Carlo - MC) datasets, referred to as $\rmD^{\rmT}_\varepsilon$, $\rmD^{\rmV}_\varepsilon$ and $\rmD^{\rm MC}_\varepsilon$, respectively, are generated. The latter are not obtained using CS, but using a uniformly random set of directions in strain space. They are mainly used for unbiased testing of the surrogate independent of the proximity to the training and validation set. The output of interest in the present work is, primarily, the effective stress, but also some error measures for the derived surrogates, which will be defined in the following sections.

Technically, the process of generating the data samples is challenging. In order to obtain reliable data, the FOM and the ROM must be evaluated thousands of times in order to obtain the needed data. Each sample consists of an effective strain~$\bueps$ and the related effective stress~$\busig$. In order to boost the performance of the simulations, a ROM-preconditioned solver for the FOM has been developed: First, an accurate (i.e. high-dimensional) ROM is solved for each load path. Then the FEM is accelerated by taking the ROM solution as initial guess for the nodal displacements during the first increment and, during the subsequent load steps, by taking the ROM displacement increment as initial guess for the FEM displacement adjustment. This not only brings the initial guess close to the final solution but it also leads to an accurate global stiffness matrix that can be combined with Quasi-Newton techniques. The ROM-accelerated FE showed a 20\% reduction in the number of Newton iterations, despite the use of a Quasi-Newton scheme. This is remarkable in view of the less accurate stiffness matrix of Quasi-Newton scheme and the faster convergence must be attributed to the improved initial guess for the FE displacement vector reconstructed from the ROM solution. Overall, this approach provides significant computational improvements over a naive FE based data generation. Further, it is noteworthy that the high-dimensional ROM solution can be used to derive a hierarchy of lower-dimensional ROM solutions needing virtually no additional Newton-Raphson iterations via linearization. More precisely the trailing entries of a ROM solution can be eliminated by making use of the Schur complement which leads to an adjustment of the remaining reduced coefficients. In our tests this downscaling of high quality ROM solutions to $N$-dimensional ROMs proved an efficient tool.

\subsubsection{Surrogate model for the effective stress}
\label{sec_stress}

\paragraph{Feature design}
\label{sec_ann_stress_feature_design}

For the successful training of ANNs the normalization of the input and output data and the design of appropriate inputs (usually referred to as features) through linear or nonlinear transformations is essential. Compared to image data and convolutional neural networks, which usually take advantage of the intrinsic connection of image data and convolution, the present input data (strain data) is low-dimensional and necessarily requires sensible mechanical guidance during feature design. From a pure data-driven perspective, general batch normalization can greatly improve the prediction quality of a network. But in the present problem setting the input and output data have a clear physical nature. Therefore, based on mechanical reasoning, the consideration of the dependency of the material law on the spherical ($\bar{\varepsilon}^\circ$) and deviatoric ($\bueps'$) degrees of freedom of the strain offers a material theoretic starting point. This linear transformation is addressed as
\begin{equation} 
	\uT^\sd(\bueps) 
	= \begin{bmatrix}
	\bar\eps^\circ \\
	\bueps'
	\end{bmatrix}
	= [\bar\eps^\circ,\bar\eps'_1,\bar\eps'_2,\bar\eps'_3,\bar\eps'_4,\bar\eps'_5]^\sfT
	\in \reals^6
	\ .
\end{equation}
Additionally, the deviatoric part of the strain can be split into its norm and direction
\begin{equation}
	\uT^\sdd(\bueps)
	= \begin{bmatrix}
	\bar\eps^\circ \\
	\norm{\bueps'} \\
	\bueps'/\norm{\bueps'}
	\end{bmatrix}
	\in \reals^7
	\ .
\end{equation}
After either of these transformations, a corresponding normalization is performed in order to prepare the strain features for the subsequent evaluation through the ANN: For $\uT^\sd$, each component of the vector $\uT^\sd(\bueps)$ is shifted and then divided by its corresponding mean and standard deviation over the training dataset $\rmD^\rmT_\varepsilon$, i.e. component-wise shifting and scaling are applied. For $\uT^\sdd$, the first component (i.e. the volumetric strain) is scaled according to the standard procedure while the deviatoric strain amplitude is divided by its peak value and the deviatoric direction remains unchanged. In the following the shifted and scaled inputs are referred to as $\ul{x}^{[0]} \in \reals^D, D=6,7$.

\paragraph{Architecture of the artificial neural network}
\label{sec_architecture}
In the present work, feedforward neural networks are used. This choice within the plethora of available artificial neural networks is driven by the fact that a function is to be calibrated that depends exclusively on the current state~$\bueps$: the effective stress of the FOM~$\busig^\mathrm{F}(\bueps)$. It should be remarked that for problems with history dependency, e.g., path-dependent plasticity or damage in cyclic loading, feedforward neural networks could, in principle, be considered, but recurrent neural networks offer much better alternatives. They are specially designed for time series and they feed back outputs of the model into the prediction of the subsequent cycle. Generally, the training costs of recurrent neural networks are immensely higher than that of feedforward neural networks, since a large number of input paths is required, instead of points in the input space. For the problem at hand recurrent neural networks offer no advantages. Hence, we choose feedforward neural networks for the rest of the present work. Hereby, networks consisting of $L>1$ layers are taken into account. For each layer~$l=1,\dots,L$ consisting of $n^{[l]}$ neurons the inputs~$\ul{x}^{[l-1]} \in \reals^{n^{[l-1]}}$ and outputs $\ul{x}^{[l]} \in \reals^{n^{[l]}}$ are related by weights $\uuW^{[l]}$, biases $\ub^{[l]}$ and activation functions~$a^{[l]}$ via the recursion
\begin{equation}
	\ux^{[l]} = a^{[l]}(\uuW^{[l]}\ux^{[l-1]} + \ub^{[l]}) \in \reals^{n^{[l]}}, \qquad
	\uuW^{[l]} \in \ffR^{n^{[l]} \times n^{[l-1]}}, \qquad
	\ub^{[l]} \in \ffR^{ n^{[l]}} \, ,
	\label{eq_ann_rec}
\end{equation}
complemented by $n^{[0]}=D$. The weights and biases of the ANN are parameters, which need to be calibrated with training data by solving an unconstrained optimization problem. The choice of activation functions is an abstract parameter that can heavily influence the quality of the surrogate. Its selection depends on the intuition of the user, complemented by thorough testing in terms of architecture sweeps. In the present context, the differentiability of the stress surrogate is aspired, as it allows for a computation of the tangent stiffness at low computational expense through automatic differentiation. This requirement naturally favours smooth activation functions. Our ANN implementation is based on Python3 (v3.4.3) using Google's TensorFlow library (v1.12.0), which offers automatic differentiation capabilities. For architecture tests the following activation functions have been used:
\begin{itemize}
\item the identity function (Id) \quad $a(x) = x$,
\item the rectified linear unit (RELU) \quad $a(x) = \max(x,0)$,
\item the softplus function (SP) \quad $a(x) = \log(1+\exp(x))$
\item and the hyperbolic tangent (TANH) \quad $a(x) = \tanh(x)$.
\end{itemize}
The identity function (Id) allows to pass unaltered input, such that a linear combination of the activation functions of the previous layer is returned. This is particularly desired in the last layer, in order to obtained an optimized linear combination of nonlinear functions as final output $\uy=\ux^{[L]}$ of the ANN. The evaluation of a single input strain through the whole ANN is addressed by the composition of all layers
\begin{equation}
	\rmann(\bueps) = \uy(\bueps) = a^{[L]} \lb \uuW^{[L]}a^{[L-1]}( \dots )+\ub^{[L]}  \rb \ .
\end{equation}

\paragraph{Loss function}

The training of the ANN requires an objective function that provides an error respecting the nature of the outputs. In the context of ANNs, the objective function is referred to as loss function. Similar to the inputs, the outputs, the effective stress of the FOM $\busig^\mathrm{F}$ defined in \eqref{eq:sigF:def}, should also be scaled using an invertible transformation
\begin{equation}
	\up(\bueps) = \uT_\sigma(\busig^{\mathrm{F}}(\bueps)) \in \reals^{d_\sigma}
	\ .
\end{equation}
Here, the same transformations $\uT^\sd$ and $\uT^\sdd$ as for the inputs are considered for $\uT_\sigma$ during architecture testing. The evaluation of the ANN is analogously abbreviated as
\begin{equation}
	\tilde\up(\bueps) = \rmann(\bueps) \in \reals^{d_\sigma}
	\ .
\end{equation}
In this work, the mean squared error (MSE) is chosen as the loss function
\begin{equation}
	\mse
	= \frac{1}{d_\sigma} \underset{\rmD^\rmT_\varepsilon}{\rmmean} \lb \nf{\up  - \tilde\up}^2 \rb
	\ .
	\label{eq_mse_sig}
\end{equation}
The MSE \eqref{eq_mse_sig} is then optimized with respect to the ANN parameters, i.e., the weights and biases are identified starting from a random initialization. The ANN output is then obtained through an inverse transformation 
\begin{equation}
	\busig^\rmann(\bueps) = \uT^{-1}_\sigma(\rmann(\bueps)) \ .
\end{equation}
It should be remarked that, from the perspective of physics-informed artificial neural networks, one may also consider the incorporation of the norm of the nonsymmetric part of the gradient $\partial \busig^\mathrm{ANN}/ \partial \bueps$ in the loss function. This would help to calibrate the network, such that its gradient is likely to be close to symmetric. But since this can not be assured for arbitrary input $\bueps$, number of layers, neurons and activation functions, the present work prefers to solely consider \eqref{eq_mse_sig} for the loss function, calibrate $\busig^\mathrm{ANN}$ as good as possible and simply symmetrize the resulting gradient $\partial \busig^\mathrm{ANN}/ \partial \bueps$. Hereby, it should be stressed that a symmetric gradient $\partial \busig^\mathrm{ANN}/ \partial \bueps$ is essential for the hyperelastic/pseudo-plastic material considered in this work, since the assembled system matrix of the macroscopic problems is symmetric by the corresponding material theory. A nonsymmetric system matrix in the macroscopic problem would also increase the computational costs, due to the thereby induced necessity for solvers for nonsymmetric matrices.

The quality of the ANN during training is checked, not with respect to the training dataset, but with the validation dataset $\rmD^\rmV_\varepsilon$ via the mean relative norm error (MRNE)
\begin{equation}
	\mathrm{MRNE}
	= \underset{\rmD^\rmV_\varepsilon}{\rmmean} \left(
	\dfrac{\nf{\busig^\rmF - \busig^\mathrm{ANN}}}{\nf{\busig^\rmF}}
	\right) .
\end{equation}
In addition to that, the mean coefficient of determination~$R^2_\sigma$ of the effective stress is evaluated
\begin{equation}
	R^2_\sigma = \frac{1}{6} \sum_{i=1}^6 R^2_i
	\ , \quad
	R^2_i  
	= 1 - 
	\dfrac{\underset{\rmD^\rmV_\varepsilon}{\rmmean} \lb \lb \bar\sigma_i^\rmF - \bar{\sigma}_i^\mathrm{ANN} \rb^2 \rb }
	{\underset{\rmD^\rmV_\varepsilon}{\rmmean} \lb ( \bar\sigma_i^\rmF )^2 \rb - \lb \underset{\rmD^\rmV_\varepsilon}{\rmmean} ( \bar{\sigma}_i^\mathrm{F} ) \rb^2}
	\ .
	\label{eq_R2}
\end{equation}
The coefficient of determination is bounded by one which is attained if and only if the surrogate coincides with the reference for all queries.

\subsubsection{Surrogate model for the error in the quantity of interest}
\label{sec_error_surrogate}

\paragraph{Error regression and classification}

In this section, we are interested in the calibration of ANNs taking strain data as input and delivering quantitative and qualitative error estimates for the stress. On the one hand, for a given strain, it might be of interest to predict the error of stress surrogate against the FOM stress. On the other hand, it might not be of particular interest to know the exact error value, but rather to know if the error is acceptable, i.e. if it is smaller than a prescribed tolerance. The quantitative error prediction leads to a classical regression problem, whereas the binarized response gives rise to an ordinary classification problem.

In the error regression problem, for a given model $\busig^\rmM \in \{\busig^{\rmR N}, \busig^\mathrm{ANN}\}$ of the effective stress, we are interested in the absolute and relative norm errors
\begin{equation}
	e_a^{\rmM}(\bueps) = \nf{\busig^\rmF(\bueps) - \busig^{\rmM}(\bueps)}
	\ , \quad
	e_r^{\rmM}(\bueps) = \frac{\nf{\busig^\rmF(\bueps) - \busig^{\rmM}(\bueps)}}{\nf{\busig^\rmF(\bueps)}}
	\ .
\end{equation}

For the error classification problem, we consider the indicator function
\begin{equation}
	\chi^\rmM(\bueps) 
	=
	\begin{cases}
	1 & 
		\text{if }e_a^\rmM(\bueps) < \tau_a 
		\ \text{or} \
		e_r^\rmM(\bueps) < \tau_r
		\\
	0 & \text{else} \ , 
	\end{cases}
	\label{eq_chi_error}
\end{equation}
with prescribed absolute and relatives tolerances $\tau_a$ and $\tau_r$, respectively. The outcome of $\chi^\rmM$ is particularly useful in order to decide on the subsequent treatment: For $\chi^\rmM=1$, the error is considered acceptable and the surrogate can be used, while $\chi^\rmM=0$ should trigger an adaptive refinement. For instance, the classifier~$\chi^\rmM$  can decide if the stress surrogate~$\busig^\rmM$ at a macroscopic integration point is acceptable or whether a more dedicated surrogate is needed.

For error regression and classification, the fully connected feed forward ANNs as described by~\eqref{eq_ann_rec} and the same activation functions as in Section~\ref{sec_architecture} are used. For the binary classification the final ANN layer is regarded as a $\log$-probability with a single neuron. This setup is usually referred to as \texttt{logits} in binary classification.

\paragraph{Loss function}

One of the desired properties, considering possible safety requirements in the error regression and classification, is to obtain if not accurate, then at least conservative results. In order to achieve a conservative behavior, for the error regression problem we consider the function
\begin{equation}
	\phi_\alpha(x) = \max(x,0) + \alpha \max(-x,0)\ ,
	\label{eq_phi_alpha}
\end{equation}
which changes the slope for negative input values to $\alpha$. The function $\phi_\alpha$ can be used to penalize underestimation of the error (for $\alpha>1$) when applied to the scalar argument of the MSE for the true error~$e^\rmM$ (representing the absolute error $e_a^\mathrm{M}$ or the relative error $e_r^\mathrm{M}$ of the model $\rmM \in \{\rmR N , \mathrm{ANN}\}$) and its ANN surrogate~$\tilde{e}^\rmM$
\begin{equation}
	\mathrm{MSE}_{\alpha} 
	= \underset{\rmD^\rmT_\varepsilon}{\rmmean} (|\phi_\alpha(e^\rM(\bueps) - \tilde{e}^\rmM(\bueps))|^2) .
\end{equation}
The MSE$_\alpha$ is considered as the loss function for error regression, where $\alpha$ acts as a penalty parameter. The corresponding $R^2$ value and the relative conservative amount (RCA) over the validation dataset
\begin{equation}
	R^2_e 
	= 1 - 
	\dfrac{\displaystyle \underset{ \rmD^\rmV_\varepsilon }{\rmmean} \lb ( e^\rmM - \tilde{e}^\rmM )^2 \rb}
	{\displaystyle \underset{ \rmD^\rmV_\varepsilon }{\rmmean} ( (e^\rmM)^2 )  - \lb \underset{ \rmD^\rmV_\varepsilon }{\rmmean}(e^\rmM) \rb^2 } \ , \qquad
	\rmrca_e = \frac{\#(\D^\rmV_\varepsilon: e^\rmM(\bueps) \leq \tilde{e}^\rmM(\bueps))}{\#(\D^\rmV_\varepsilon)}
\end{equation} 
are used to assess the quality of the prediction.

For the error classification of model $\rmM \in \{\rmR N, \mathrm{ANN}\}$, due to the binary nature of \eqref{eq_chi_error}, the last layer of the ANN is defined as the composition of a standard sigmoid function and a shifted step function, i.e., 
\begin{equation}
    \tilde{\chi}^\rmM(\bueps) = s \circ \tilde{\chi}^\rmM_0(\bueps), \qquad
    \tilde\chi^\rmM_0(\bueps) = \frac{1}{1+\exp(-\mathrm{ANN}(\bueps))} , \qquad
	s(x)
	= \begin{cases}
	1 & x > 1/2, \\
	0 & \text{else.}
	\end{cases}
\end{equation}
The loss function for classification chosen in this work is the weighted binary cross entropy
\begin{equation}
	\eta_{w} 
	= -\underset{\rmD^\rmT_\varepsilon}{\rmmean} \, \big(
	w \ \chi^\rmM(\bueps) \, \log\lb\tilde\chi^\rmM_0(\bueps)\rb
	+ (1-\chi^\rmM(\bueps)) \, \log\lb1-\tilde\chi^\rmM_0(\bueps)\rb
	 \big)
	\ .
\end{equation}
Herein, false positive predictions dominate the cross entropy for~$w>1$, while $0<w<1$ puts the focus on false negative classification. We define the overall accuracy of the classifier as the expectation of finding the same response in the true indicator~$\chi^\rmM$ and in the surrogate~$\tilde{\chi}^\rmM$:
\begin{align}
 \rmacc = 1 - \underset{\rmD^\rmV_\varepsilon}{\rmmean} \, \lb \vert \tilde{\chi}^\rmM - \chi^\rmM \vert  \rb .
\end{align}
Further, the accuracy within the bin $b \in \{ 0, 1 \}$ is defined as the conditional probability
\begin{equation}
	\rmacc_{b} = 1 - \underset{ \{ \ux \in \rmD^\rmV_\varepsilon: \chi^\rmM(\ux) = b \} }{\rmmean} \, \lb \vert \tilde{\chi}^\rmM - \chi^\rmM \vert  \rb .
\end{equation}
The reader should note, that $\rmacc_0$ is more relevant when seeking conservative estimates. Only if $\rmacc_0$ and $\rmacc_1$ are close to unity, then the overall classification is robust, while for seemingly good $\rmacc$ (e.g. around 0.98) the critical $\rmacc_0$ could be inappropriate. This effect is particularly important if the surrogate has only few outliers requiring further processing.

% FF+MF
\subsection{Hybrid ANN/ROM multi-level Finite Element simulation}
\label{sec:Adaptive}
\subsubsection{General hybrid approach}
In order to build a twoscale simulation model relying on the finite element method on the larger scale, the material model must be replaced by the homogenized response of the heterogeneous solid. In Sections~\ref{sec:ROM} and \ref{sec_stress} the use of ROM and ANN serving as surrogates for the effective stress tensor and the effective tangent stiffness are described in detail. Both surrogates can be combined by introducing an indicator function~$\chi(\ol{\fx}): \ol{\varOmega} \mapsto \{0; 1\} $ which adaptively selects between the rapid and purely data-driven (but less physical) ANN if $\chi=1$ and the physics-driven ROM for $\chi=0$. The indicator function represents the binarized confidence in the accuracy of the ANN surrogate.

First, a simple ansatz for $\chi$ is chosen by setting $\chi$ to one if the current strain at the macroscopic position~$\ol{\fx}\in\ol{\varOmega}$ falls within the region covered by samples during the training of the ANN. In the present study this is equivalent to the kinematic indicator
\begin{align}
 \chi^{\rm K}(\ol{\fx}) 
 	&= \left\lbrace 
 	\begin{array}{rl} 
 	1 & \text{if } \Vert\fbeps(\ol{\fx})\Vert_W \leq \ol{\varepsilon}_0,\\ 
 	0 & \text{else}.
 	\end{array} \right.
\end{align}
Here, $\ol{\varepsilon}_0={\rm max}(\rmD_r)$ is the peak amplitude used during Concentric Sampling and $\Vert \cdot \Vert_W$ denotes a weighted norm that transforms elements of $\rmD_\varepsilon$ defined via \eqref{eq_dataset_strain} back into normalized directions:
\begin{align}
 \Vert \ol{\ueps} \Vert_W &= \sqrt{ \hat{r}^2 \, \Vert {\rm sph}(\ol{\ueps}) \Vert^2_2 + \Vert {\rm dev}(\ol{\ueps}) \Vert^2_2 } .
\end{align}
The use of the ROM outside of the training domain is motivated by its reluctance to energy minimization, i.e. by preserving the key physical characteristics of the full order model while restricted to a relevant subspace of the solution manifold.

A second indicator can be obtained by evaluating the accuracy of the ANN. Therefore, a binary classifier~$\tilde\chi^\mathrm{ANN}: Sym(\reals^{3 \times 3}) \mapsto \{ 0, 1 \}$ is employed following the procedure outlined in Section~\ref{sec_error_surrogate}. The indicator function is then replaced by the classifier: $\chi(\ol{\fx}) = \tilde\chi^\mathrm{ANN}( \fbeps(\ol{\fx}) )$.

\subsubsection{Technical issues related to on-the-fly model switching}

\begin{figure}[h]
\centering
\includegraphics[scale=0.9]{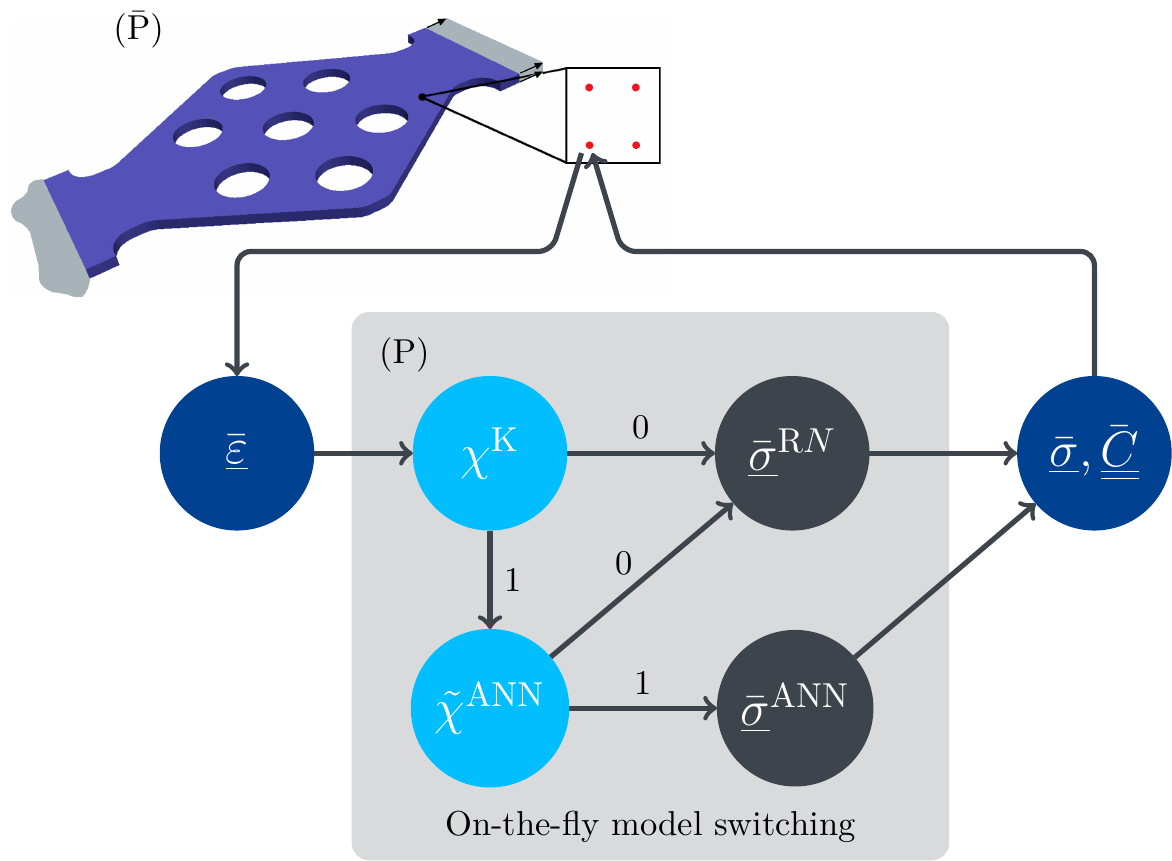}
\caption{Macroscopic FE boundary value problem $(\bar{\mathrm{P}})$ with on-the-fly model switching at integration points for the computation of the effective microscopic stress $\busig$ for prescribed microscopic effective strain $\bueps$ for the microscopic problem $(\mathrm{P})$: first, $\chi^\mathrm{K}$ checks if $\bueps$ is in the training region of the ANN surrogate for $\busig$; if the quality of ANN is acceptable based on $\tilde\chi^\mathrm{ANN}$, then $\busig^\mathrm{ANN}$ is evaluated and passed to the macroscopic FE problem; but if either $\bueps$ is outside of the ANN training range or the ANN surrogate is not accurate enough, then a previously selected accurate ROM of corresponding dimension $N$ is evaluated and then passed to the macroscopic problem.}
\label{fig_flowchart}
\end{figure}

At first, the concept of the indicator function~$\chi$ marking the confidence region for the ANN and employing the ROM elsewhere sounds straight-forward. However, this simple approach does not work in practice as the two concurrent surrogates do not provide continuous approximations of the stresses. This can be illustrated by letting $\cC \subseteq Sym( \ffR^{3\times 3} )$ denote the confidence region of the ANN in strain space. It should be noted, that $\cC$ may contain several holes depending on the chosen quality indicator determining a point or region in strain space as admissible or not. On the boundary~$\partial\cC$ of the confidence region there is a hard transition between the two surrogates which induces a stress jump, leading to a non-smooth material response. When switching between ANN and ROM on-the-fly, i.e., when deciding for each query adaptively which surrogate should be evaluated, convergence of the macroscopic problem is disrupted, rendering the straight-forward implementation of a quality indicator guided adaptive procedure infeasible. One may try to solve this problem with multi-fidelity approaches, see, e.g., \cite{Meng2019}, where multiple nested surrogates (e.g., artificial neural networks) based on data groups of different accuracy/fidelity and amounts are trained. Unfortunately, such multi-fidelity data approaches are not applicable for the problem at hand. In order to motivate this more clearly, consider again Fig.~\ref{fig_fe2} and the strategy illustrated in Fig.~\ref{fig_flowchart} for a macroscopic boundary value problem solved with FE and calling for an on-the-fly model switching at the integration points for the computation of $\busig$ for prescribed $\bueps$.

In the context of twoscale simulations, the problem is not the accuracy/fidelity of the training data of the microscopic problem, but (1) the usage of a surrogate outside of its training range (based on $\chi^\mathrm{K}$ for the ANN stress surrogate) and (2) the point-wise quality of the surrogate with respect to prescribed tolerances ($\tilde\chi^{\mathrm{ANN}}$ for the ANN effective stress surrogate), which define the boundary of the confidence region $\cC$ and trigger the model switching. Both events can occur in twoscale simulations, since the input field at the macroscopic scale (i.e., $\bar\feps(\bar\fx)$) is not known for arbitrary macroscopic geometry and boundary conditions, such that point-wise at the macroscopic scale the ANN microscopic surrogate for the effective stress may be evaluated far outside of its training range or may be inaccurate. If the ANN effective stress surrogate is inaccurate, then, e.g,  a fixed ROM of sufficient accuracy can be initiated, as depicted in Fig.~\ref{fig_flowchart}. Naturally, in order to lower the number of ROM evaluations, one could simply enhance the existing networks $\busig^\mathrm{ANN}$ and $\tilde\chi^\mathrm{ANN}$ during the online computation by re-training using additional samples. However, there is no methodology available that can a priori guaranty accuracy gains without the need of extensive architecture sweeps and substantial sampling of extended and/or refined regions in the input space. Therefore, such an online re-training is not a viable option at the moment and alternatives need to be investigated. Contrary to the inherent properties of ANNs and the related training, (i) the ROM solution is obtained in a physically guided procedure, (ii) the errors of the ROMs drop with increasing dimension and (iii) the ROM has no intrinsic validity domain limitation in strain space. This motivates the use of a ROM of sufficient dimension outside of the validity domain of the ANN stress surrogate. Approaches for the algorithmic realization of the dynamic switching between concurrent surrogates are described in the sequel.

\paragraph{Staggered hybrid ANN/ROM algorithm}

The first approach consists of a staggered procedure, where the ANN is used as the only stress surrogate in a first run of the twoscale simulation (see~Algorithm~\ref{algo:fe2r-nn:a}). Thereby, a first overall response is gathered. This is followed by a second run, in which the subset of all integration points having seen a zero quality indicator during any of the load steps of the first run are enforced to use the ROM surrogate. This set is then kept constant, i.e. switching from ANN to ROM is one way. This procedure enables the use of the ANN solution as an initial guess for the subsequent hybrid run which leads to low iteration counts and improved performance. During the second run, the difference of the ANN and the ROM can be evaluated to provide valuable post-processing data in order to better understand the quantitative impact of the model modifications, see also examples in Section~\ref{sec:fe2r:algo:a}. Two major disadvantages of this approach are (i) the irreversibility of the ROM activation which can lead to substantial computational costs and (ii) the possible failure during the first run, if the ANN surrogate becomes nonconvergent. The latter can, e. g., occur if the local magnitude of~$\fbeps$ on the macroscale falls way outside of range of the training data.

\begin{algorithm}[h]
\small
	%%%%%%%%%%%%%%%%%%%%%%%%%%%%%%%%%%%%%%%%%%%%%%%%%%%%%%%%%%%%%%%%%%%%%%%%%%%%%%%%%%%%%%%%%%%%%%%
	\SetAlgoLined
	\SetEndCharOfAlgoLine{ }
	%%%%%%%%%%%%%%%%%%%%%%%%%%%%%%%%%%%%%%%%%%%%%%%%%%%%%%%%%%%%%%%%%%%%%%%%%%%%%%%%%%%%%%%%%%%%%%%
	\Input{quality indicator~$q$; ANN surrogate $\ol{\usigma}^{\rm ANN}$; ROM surrogate $\ol{\usigma}^\rn$
	%\\{\it [optional]} quality indicator~$\chi^{\rmC}$
    }
	%%%%%%%%%%%%%%%%%%%%%%%%%%%%%%%%%%%%%%%%%%%%%%%%%%%%%%%%%%%%%%%%%%%%%%%%%%%%%%%%%%%%%%%%%%%%%%%
    \lFor(\tcp*[f]{initialize quality indicator}){$i=1,\dots,n_{\rm gp}$}{$q_i=1$;}
	\For{$i_{\rm inc}=1, \dots, n_{\rm inc}$}{
        \While{increment not converged}{
        Newton-Raphson iteration using ANN surrogate only: $\ol{\usigma}^{\rm ANN}$\\
	\lFor(\tcp*[f]{update quality indicator (one way update)}){$i=1, \dots, n$}{$q_i= {\rm min}(q_i, \chi(\fbeps_i))$}
        }
        converged nodal displacements $\to \ul{u}^{(1)}(i_{\rm inc})$ \tcp*{level 1 solution}
    }
	%%%%%%%%%%%%%%%%%%%%%%%%%%%%%%%%%%%%%%%%%%%%%%%%%%%%%%%%%%%%%%%%%%%%%%%%%%%%%%%%%%%%%%%%%%%%%%%
	{\bf restart simulation} (conserve quality indicators~$q_i$, $i=1,\dots,n_{\rm gp}$)\tcp*{second run}
	\For{$i_{\rm inc}=1, \dots, n_{\rm inc}$}{
        initial guess using previous simulation run: $\ul{\Delta u} = \ul{u}^{(1)}(i_{\rm inc}) - \ul{u}^{(1)}(i_{\rm inc}-1)$\\
        \While{increment not converged}{
        Newton-Raphson iteration using ANN (if $q_i=1$) or ROM (if $q_i=0$)
%         {\it [optional]} select tangent operator of the ANN and ROM cf. to a second indicator function~$\chi^{\rmC}$\\
        }
        converged nodal displacements $\to \ul{u}^{(2)}(i_{\rm inc})$ \tcp*{level 2 solution}
    }
	%%%%%%%%%%%%%%%%%%%%%%%%%%%%%%%%%%%%%%%%%%%%%%%%%%%%%%%%%%%%%%%%%%%%%%%%%%%%%%%%%%%%%%%%%%%%%%%
	%%%%%%%%%%%%%%%%%%%%%%%%%%%%%%%%%%%%%%%%%%%%%%%%%%%%%%%%%%%%%%%%%%%%%%%%%%%%%%%%%%%%%%%%%%%%%%%
	\caption{Staggered hybrid ANN/ROM twoscale simulation algorithm}
 \label{algo:fe2r-nn:a}
\end{algorithm}

\paragraph{Adaptive on-the-fly ANN/ROM algorithm}
A second on-the-fly model selection procedure, solving both of the aforementioned issues, is described in Algorithm~\ref{algo:fe2r-nn:b}: It re-initializes the quality indicator in favor of the ANN at the beginning of each load increment. During the subsequent nonlinear Newton-Raphson iterations of the same increment, the indicator is updated in a monotonic way, i.e. switching from ANN to ROM is allowed but not vice verse (see line~\ref{q-update} in Algorithm~\ref{algo:fe2r-nn:b}). The computational efficiency can be improved by substituting only part of the equilibrium iteration by the ROM.

\begin{algorithm}[h]
\small
	%%%%%%%%%%%%%%%%%%%%%%%%%%%%%%%%%%%%%%%%%%%%%%%%%%%%%%%%%%%%%%%%%%%%%%%%%%%%%%%%%%%%%%%%%%%%%%%
	\SetAlgoLined
	\SetEndCharOfAlgoLine{ }
	%%%%%%%%%%%%%%%%%%%%%%%%%%%%%%%%%%%%%%%%%%%%%%%%%%%%%%%%%%%%%%%%%%%%%%%%%%%%%%%%%%%%%%%%%%%%%%%
	\Input{quality indicator~$q$; ANN surrogate $\ol{\usigma}^{\rm ANN}$; ROM surrogate $\ol{\usigma}^\rn$;
    %\\{\it [optional]} quality indicator~$\chi^{\rmC}$
    }
	%%%%%%%%%%%%%%%%%%%%%%%%%%%%%%%%%%%%%%%%%%%%%%%%%%%%%%%%%%%%%%%%%%%%%%%%%%%%%%%%%%%%%%%%%%%%%%%
	\For{$i_{\rm inc}=1, \dots, n_{\rm inc}$}{
        \lFor(\tcp*[f]{reset quality indicator}){$i=1,\dots,n_{\rm gp}$}{$q_i=1$;}
%         \tcp*{reset indicator at the beginning of each load increment}
        \While{increment not converged}{
            \For{$i=1, \dots, n_{\rm gp}$}{
                evaluate strain~$\fbeps_i$ and update quality indicator $q_i= {\rm min}(q_i, \chi(\fbeps_i))$ \tcp*{one way update} \label{q-update}
                use $\ol{\usigma}^{\rm ANN}$ if $q_i=1$ and $\ol{\usigma}^\rn$ if $q_i=0$\\
                {\it[optional]} compute difference of the stress between $\ol{\usigma}^{\rm ANN}$ and $\ol{\usigma}^\rn$ \tcp*{post-processing}
            }
        }
        converged nodal displacements $\to \ul{u}(i_{\rm inc})$ 
    }
	%%%%%%%%%%%%%%%%%%%%%%%%%%%%%%%%%%%%%%%%%%%%%%%%%%%%%%%%%%%%%%%%%%%%%%%%%%%%%%%%%%%%%%%%%%%%%%%
    \caption{Adaptive on-the-fly ANN/ROM twoscale simulation algorithm}
 \label{algo:fe2r-nn:b}
\end{algorithm}

\section{Numerical examples}
\label{sec_nex}

%FF
\subsection{Underlying material model}
\label{sec_micro_mat}
\begin{figure}[h]
 \centering
 \includegraphics[height=50mm]{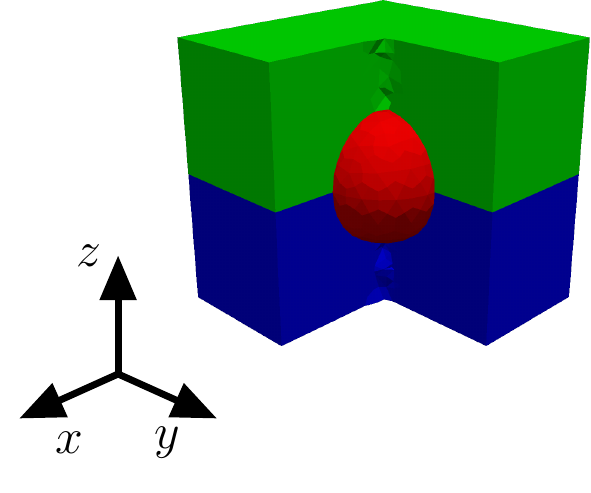} \qquad
 \includegraphics[height=50mm]{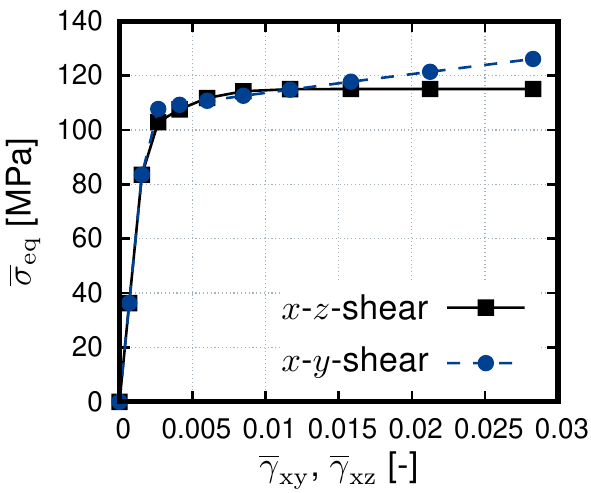} \qquad
 \caption{{\it left:} Periodic representative volume element (RVE) of the triphasic material: layer 1 (blue; pseudo-plastic with hardening), layer 2 (green; pseudo-plastic, no hardening) and inclusion (red; linear elastic);
 {\it right:} directional dependency of the material response}
 \label{fig:rve}
\end{figure}

An artificial heterogeneous solid consisting of three phases is investigated. It consists of a laminate structure of two pseudo-plastic materials where the two layers share the same elastic parameters ($E_1=E_2=75$~GPa, $\nu_1=\nu_2=0.3$) but have different yield strength and hardening behavior: The first layer has a yield stress of 100~MPa and a linear hardening slope of 2000~MPa, whereas the second layer has a yield stress of 115~MPa in the absence of hardening. The third phase is represented by a spherical inclusion that is centered on the interface of the two phases. The inclusion is assumed linear elastic with properties mimicking a ceramic inclusion made of SiC ($E=400$~GPa, $\nu=0.2$), see Fig.~\ref{fig:rve}. The volume fractions of the two plastic layers are 46.73\% each and the one of the inclusion is 6.54\%. The material was designed to induce a directional dependency of the effective material behavior (see right plot in Figure~\ref{fig:rve} for an example). This feature makes the identification of the unknown homogenized response more challenging and, thereby, a benchmark problem for the developed methodology is designed.

%MF
\subsection{Quantitative comparison of ROM and ANN surrogate models}\label{sec:ANNvsROM}
\subsubsection{Effective stress surrogate}

The strain space is sampled as described in Sect.~\ref{sec:data:generation} for an effective strain amplitude discretization $\rmD_r = \{0.0005,0.002,0.0035,0.005,0.0075,0.01,0.015,0.025,0.04\}$. The spherical / volumetric part of the primal strain dataset is rescaled with $\hat{r} = 5$. Then, 1152 training, 288 validation and 512 Monte Carlo directions are generated, yielding 10368 training, 2592 validation and 4608 Monte Carlo effective strain points in $\reals^6$. 

An initial architecture testing phase is conducted. The activation functions and transformations illustrated in Sect.~\ref{sec_stress} are considered, together with varying number of layers and neurons. The architecture test with $L \in \{3,\dots,6\}$ and number of neurons per hidden layer $n^{[l]} \in \{16,32,64,128\}, l \in \{1,\dots,L-1\}$, yields that none of the activation functions (RELU, SP, TANH) show a remarkable advantage over the other, even for as large number of epochs as 10,000 with whole batch training for a learning rate of 0.001 using an ADAM optimizer. However, the feature design of input (effective strain) and output data (effective stress of the FOM) has a major influence. Hereby, the most successful combination is identified to be the use of the spherical-deviatoric transformation $\uT^\sd$ for the input as well as for the output. The transformation $\uT^\sdd$ did not show major advantages in the final objective function values.

\begin{figure}[h]
\includegraphics[width=0.95\textwidth]{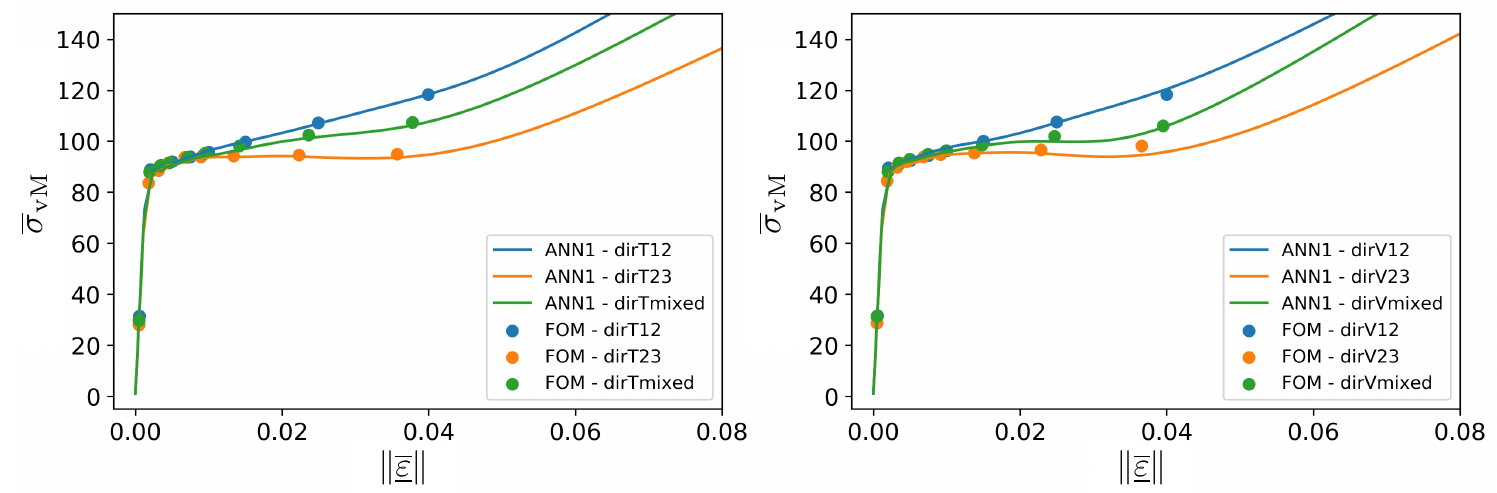}
\caption{Von Mises effective stress versus effective strain norm for ANN1 for the 3 loading directions of the training dataset ({\it left}) and 3 loading directions of the validation dataset ({\it right}) tabulate in Tab.~\ref{tab_dirs_vM}}
\label{fig_sigma_ann1_vm}
\end{figure}

Based on the initial architecture testing, the softplus function (SP) has been chosen to power further investigations, due to its monotonic and differentiability properties in regard of an expected monotonic stress behavior and need for tangent operators for future FE multiscale computations. In Tab.~\ref{tab_sigma_anns}, different architectures are tabulated, showing the performance of each ANN. Based on the MRNE and $R^2_\sigma$ values for the validation dataset (and the corresponding values MRNE$_\mathrm{MC}$ and $R^2_{\sigma \mathrm{MC}}$ evaluating the MC dataset), the ANN1 comprised of six layers with five softplus hidden layers and 128 neurons per hidden layer is chosen for the final evaluation. In Fig.~\ref{fig_sigma_ann1_vm} the prediction of ANN1 for the von Mises effective stress $\bar\sigma_{vM}$ is depicted for the three in Tab.~\ref{tab_dirs_vM} tabulated directions of the training (dirT12, dirT23 and dirTmixed) and validation datasets (dirV12, dirV23 and dirVmixed), showing a good agreement with the FOM data. It should be noted that the directions dirT/V12 have a (12) dominant component, meaning that the hardening material shown in Fig.~\ref{fig:rve} is activated, while dirT/V23 have a (23) dominant component allowing for a localization of the deformation in the non-hardening material, see Fig.~\ref{fig_sigma_ann1_vm}. The effective strain directions dirT/Vmixed show some examples for combined loading and corresponding material response, see Fig.~\ref{fig_sigma_ann1_vm}. The reader should take into account, that the ANNs have been trained with strain data up to a norm of 0.04 in the primal strain set $\hat\rmD^\rmT_\varepsilon$ (corresponding to the last data point for each loading direction in Fig.~\ref{fig_sigma_ann1_vm}). The behavior of the ANN1 beyond this norm value was expected to tend to keep increasing due to the properties of the softplus function. However, due to the tendency of the ANN to increasingly overestimate the stresses and the artificial stiffening at load amplitudes beyond the training data, ANN1 is not expected to deliver accurate results beyond an effective strain norm of approximately 0.04 in respect to the primal strain set $\hat\rmD^\rmT_\varepsilon$. Finally, in addition to the a posteriori symmetrization of the gradient $\partial \busig^\mathrm{ANN1}/\partial \bueps$, numerical tests were carried out to verify that (i) the gradient obtained via automatic differentiation is almost symmetric (with an average error lower than 1\%) and (ii) that the difference of the symmetrized gradient to the algorithmic tangent of the ROM with 96 modes was matched up to relative errors around 1.5\%. These two checks approved the chosen approach. For a better transparency of these results, the authors offer \ref{supp_data}, see section \ref{supp_mat}, containing the FOM data, the trained ANN1 and commands for the reproduction of all corresponding results.

\begin{table}[h]
\begin{center}
\caption{ANNs for the effective stress surrogate with corresponding choice of input features, network architecture, intermediate transformation of stress data $\uT_\sigma$, measures MRNE and $R^2_\sigma$ for the validation dataset and MRNE$_\mathrm{MC}$ and $R^2_{\sigma \mathrm{MC}}$ for the MC dataset}
\label{tab_sigma_anns}
\begin{tabular}{l|ccc|cc|cc}
ANN ID 
	& Features
	& Architecture
	& $\uT_\sigma$
	& MRNE 
	& $R^2_\sigma$ 
	& MRNE$_\mathrm{MC}$
	& $R^2_{\sigma\mathrm{MC}}$
	\\\hline
$\busig^\mathrm{ANN1}$
	& $\uT^\sd$
	& $\{5 \times 128 (\mathrm{SP}) - 6 (\mathrm{Id})\}$
	& $\uT^\sd$
	& 0.0189
	& 0.9995
	& 0.0183
	& 0.9995
	\\
$\busig^\mathrm{ANN2}$
	& $\uT^\sd$
	& $\{5 \times 64 (\mathrm{SP}) - 6 (\mathrm{Id})\}$
	& $\uT^\sd$
	& 0.0204
	& 0.9995
	& 0.0200
	& 0.9994
	\\
$\busig^\mathrm{ANN3}$
	& $\uT^\sdd$
	& $\{5 \times 64 (\mathrm{SP}) - 6 (\mathrm{Id})\}$
	& $\uT^\sd$
	& 0.0241
	& 0.9995
	& 0.0241
	& 0.9995
	\\
$\busig^\mathrm{ANN4}$
	& $\uT^\sd$
	& $\{5 \times 16 (\mathrm{SP}) - 6 (\mathrm{Id})\}$
	& $\uT^\sd$
	& 0.1578
	& 0.9768
	& 0.1564
	& 0.9751
	\\
\end{tabular}
\end{center}
\end{table}

\begin{table}[h]
\caption{Effective strain load directions $\bueps/\nf{\bueps}$ for the inspection of the effective von Mises stress $\bar{\sigma}_\mathrm{vM}$ in the evaluation of ANN1}
\label{tab_dirs_vM}
\begin{center}
\begin{tabular}{l|ccccccc}
Direction ID & \multicolumn{7}{l}{Direction of $(\bar\varepsilon_{11},\bar\varepsilon_{22},\bar\varepsilon_{33},\sqrt{2}\bar\varepsilon_{12},\sqrt{2}\bar\varepsilon_{13},\sqrt{2}\bar\varepsilon_{23}) \in \ffR^6$} \\ \hline
dirT12 
	& (-0.10
	& -0.07
	& 0.15
	& 0.96
	& 0.11
	& 0.16)
	\\
dirT23
	& (-0.03
	& -0.10
	& -0.05
	& 0.00
	& 0.08
	& 0.99)
	\\
dirTmixed 
	& (-0.12
	& 0.03
	& -0.03
	& 0.48
	& -0.16
	& 0.85)
	\\ \hline
dirV12
	& (-0.11
	& -0.15
	& 0.27
	& 0.89
	& -0.27
	& -0.18)
	\\
dirV23
	& (-0.11
	& 0.02
	& -0.07
	& -0.12
	& -0.08
	& 0.98)
	\\
dirVmixed
	& (0.02
	& -0.31
	& 0.24
	& 0.04
	& -0.12
	& 0.91)
\end{tabular}
\end{center}
\end{table}

\subsubsection{Error surrogates}
\label{sec:ann:classifier}

For the error regression and classification, it is first necessary to gain an overview regarding the quality of the $N$-dimensional ROMs and of the best of the trained ANN effective stress surrogates $\busig^{\mathrm{ANN}1}$ of the previous section. 

\begin{figure}[h]
\includegraphics[width=0.95\textwidth]{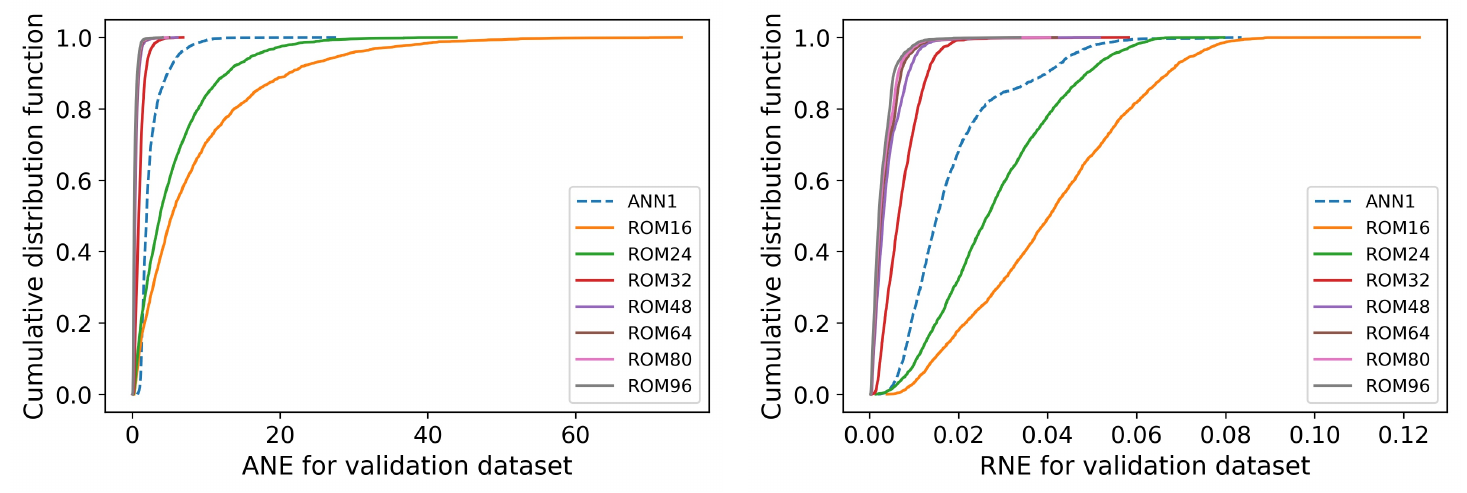}
\caption{Cumulative distribution function of the errors of ANN and ROMs of dimensions 16 to 96: distribution of ANE $e_a^\rmM$ ({\it left}); distribution of RNE $e_r^\rmM$ ({\it right})}
\label{fig_error_distributions}
\end{figure}

In Fig.~\ref{fig_error_distributions} the cumulative distribution function of the absolute norm error $e_a^\rmM$ (ANE) and of the relative norm error $e_r^\rmM$ (RNE) for the validation set $\rmD^\rmV_\varepsilon$ are shown for ROMs of different dimensions $N$ and for $\busig^\mathrm{ANN1}$. It is clearly visible that for increasing ROM dimension, the accuracy of the ROM improves for both, the ANE and RNE. This is expected, since the higher the ROM dimension, the richer the underlying function space, i.e. the distance to the solution manifold of the full order model decreases. It should be noted that the ANN effective stress model $\busig^\mathrm{ANN1}$ performs well against ROM16 and ROM24. The ROM32 yields a mean ANE of 1.019 MPa and a mean RNE of 0.007. It is from now on assumed that the accuracy of the ROM32 suffices for future multiscale FE simulations, i.e. an a priori quality assessment is made.

We first demonstrate the error regression in terms solely of the $N$-dimensional ROMs for the corresponding ANE and RNE. These error measures could be used for an adaptive selection of a ROM after having access to its estimated errors. An architecture test for ANNs with number of layers $L \in \{3,\dots,6\}$, neurons per hidden layer $n^{[l]} \in \{16,32,64\}$, up to 10,000 epochs and whole batch training is performed. A selection of the trained ANNs is tabulated in Tab.~\ref{tab_error_reg}. 

\begin{table}[h]
\begin{center}
\caption{ANNs for error regression with corresponding choice of input feature, network architecture, penalty parameter $\alpha$, corresponding quality indicators $R^2_e$ and RCA$_e$ for the validation dataset and $R^2_{e \mathrm{MC}}$ and RCA$_{e \mathrm{MC}}$ for the MC dataset}
\label{tab_error_reg}
\begin{tabular}{l|ccc|cc|cc}
ANN ID
	& Features
	& Architecture
	& $\alpha$
	& $R^2_e$
	& $\rmrca_e$
	& $R^2_{e\mathrm{MC}}$
	& $\rmrca_{e\mathrm{MC}}$
	\\ \hline
$\tilde{e}_a^\mathrm{R16|1}$
	& $\uT^\sd$
	& $\{4 \times 64(\mathrm{SP}) - 1 (\mathrm{Id})\}$
	& 3
	& 0.9868
	& 0.7924
	& 0.9892
	& 0.8082
	\\
$\tilde{e}_a^\mathrm{R16|2}$
	& $\uT^\sd$
	& $\{4 \times 64(\mathrm{RELU}) - 1 (\mathrm{Id})\}$
	& 1
	& 0.9904
	& 0.5116
	& 0.9921
	& 0.5371
	\\
$\tilde{e}_a^\mathrm{R24|1}$
	& $\uT^\sdd$
	& $\{4 \times 64 (\mathrm{TANH}) - 1 (\mathrm{Id})\}$
	& 3
	& 0.9733
	& 0.7948
	& 0.9741
	& 0.7995
	\\
$\tilde{e}_a^\mathrm{R24|2}$
	& $\uT^\sdd$
	& $\{4 \times 64 (\mathrm{TANH}) - 1 (\mathrm{Id})\}$
	& 1 
	& 0.9906
	& 0.5305
	& 0.9895
	& 0.5206
	\\ \hline
$\tilde{e}_r^\mathrm{R16|1}$
	& $\uT^\sd$
	& $\{5 \times 64 (\mathrm{RELU}) - 1 (\mathrm{Id})\}$
	& 3
	& 0.8525
	& 0.7323
	& 0.8642
	& 0.7227
	\\
$\tilde{e}_r^\mathrm{R16|2}$
	& $\uT^\sd$
	& $\{5 \times 64 (\mathrm{RELU}) - 1 (\mathrm{Id})\}$
	& 1
	& 0.9080
	& 0.5104
	& 0.9259
	& 0.4957
	\\
$\tilde{e}_r^\mathrm{R24|1}$
	& $\uT^\sd$
	& $\{5 \times 64 (\mathrm{RELU}) - 1 (\mathrm{Id})\}$
	& 3
	& 0.7822
	& 0.7562
	& 0.8316
	& 0.7574
	\\
$\tilde{e}_r^\mathrm{R24|2}$
	& $\uT^\sd$
	& $\{5 \times 64 (\mathrm{RELU}) - 1 (\mathrm{Id})\}$
	& 1
	& 0.8923
	& 0.4884
	& 0.9170
	& 0.5002
	\\
\end{tabular}
\end{center}
\end{table}

\begin{figure}[h]
\begin{center}
\includegraphics[width=0.95\textwidth]{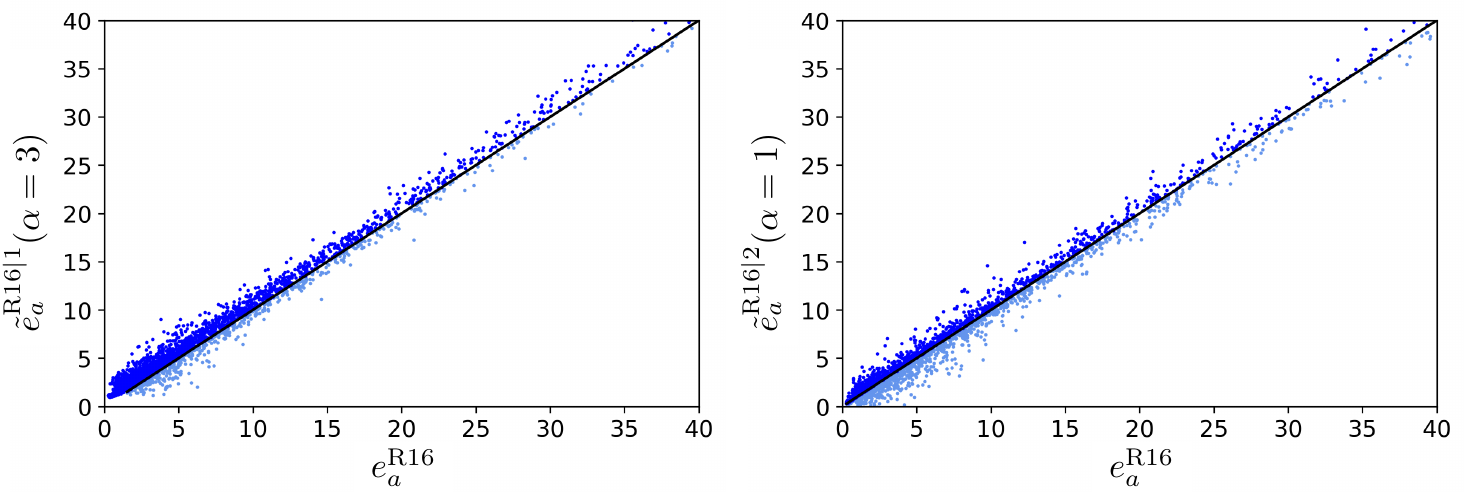}
\caption{Correlation plots for the ROM16 ANE and corresponding error regression ANNs in the range [0,40] MPa: $\tilde{e}_a^{\mathrm{R16|1}}$ with penalization of error underestimation parameter $\alpha = 3$ ({\it left}) and $\tilde{e}_a^{\mathrm{R16|2}}$ with $\alpha = 1$ ({\it right})}
\label{fig_ane_rom16}
\end{center}
\end{figure}

The ANNs $\tilde{e}_a^\mathrm{R16|1/2}$, tabulated in Tab.~\ref{tab_error_reg}, are depicted in Fig.~\ref{fig_ane_rom16}. The influence of the penalty parameter $\alpha$, introduced in \eqref{eq_phi_alpha}, can be seen in Fig.~\ref{fig_ane_rom16} (left plot), where it becomes visible that the larger amount of points are found on the upper side of the diagonal, i.e., the predicted error is larger than the error in the validation set. This is reflected in the relative conservative amount (RCA), see Tab.~\ref{tab_error_reg}. The usual trade-off is that increasing $\alpha$ yields conservative behavior (i.e., a higher RCA), but reduces the accuracy in terms of $R^2_e$. Analog behavior is observed for $e_r$, as tabulated in Tab.~\ref{tab_error_reg} (bottom half). Large values of $\alpha$ yield reduced $R^2_e$ values, due to the dilemma of balancing a reduction of the loss function, while preserving conservative behavior.  

The error classification is conducted for the absolute and relative tolerances $\tau_a = 2 \text{MPa}$ and $\tau_r = 0.02$, respectively. Architecture testing for $L \in \{3,\dots,6\}$ and $n^{[l]} \in \{16,32,64\}$ for the hidden layers yield varying quality of results depending on the weight $w$ on the false positive. Depending on the number of positive and negative outcomes, the weights should be adapted. For the architecture testing of this work, the ratio
\begin{equation}
	w_0 = \frac{\#(\rmD^\rmT_\varepsilon : \chi(\bueps) = 0)}{\#(\rmD^\rmT_\varepsilon : \chi(\bueps) = 1)}
\end{equation}
is considered. If the number of negative outcomes in the training data $\#(\rmD^\rmT_\varepsilon : \chi(\bueps) = 0)$ is higher than the positives, then $w_0>1$ holds. The consideration of $w = w_0$ in the binary cross entropy partly equilibrates the influence of the false positive (i.e. classified accurate but violating the tolerance) and false negative (i.e. classified inaccurate but within tolerance). But it may also overly bias the cross entropy during training, yielding poor accuracy in one bin. Therefore, $w$ is sampled between unity and $w_0$ in four evenly spaced steps during architecture testing. A selection of trained ANNs is tabulated in Tab.~\ref{tab_ann_class}. 

\begin{table}[h]
\begin{center}
\caption{ANNs for error classification for $\busig^{\rmR 16/24/32}$ and $\busig^{\mathrm{ANN}1}$ of previous section with tolerances $\tau_a = 2 \text{MPa}$ and $\tau_r = 0.02$}
\label{tab_ann_class}
\begin{tabular}{l|c|ccc|ccc}
ANN ID
	& $w_0$
	& Features
	& Architecture
	& $w$
	& $\mathrm{ACC}$
	& $\mathrm{ACC}_{0}$
	& $\mathrm{ACC}_{1}$
	\\\hline
$\tilde\chi^{\mathrm{R}16}$
	& 2.9184
	& $\uT^\sd$
	& $\{5 \times 64(\mathrm{TANH}) - 1(\mathrm{Id})\}$
	& 1
	& 0.9282
	& 0.9401
	& 0.8905
	\\
$\tilde\chi^{\mathrm{R}24}$
	& 1.6328
	& $\uT^\sd$
	& $\{5 \times 64(\mathrm{TANH}) - 1(\mathrm{Id})\}$
	& 1.4746
	& 0.9047
	& 0.9077
	& 0.8999
	\\
$\tilde\chi^{\mathrm{R}32}$
	& 0.0047
	& $\uT^\sdd$
	& $\{3 \times 64(\mathrm{RELU}) - 1(\mathrm{Id})\}$
	& 0.0047
	& 0.9483
	& 0.7778
	& 0.9495
	\\
$\tilde\chi^{\mathrm{ANN}1}$
%ACC0.8611_RA0_0.6721_RA1_0.8831_64_64_64_64_64_break
	& 0.1256
	& $\uT^\sdd$
	& $\{5 \times 64(\mathrm{RELU}) - 1(\mathrm{Id})\}$
	& 0.3442
	& 0.8611
	& 0.6721
	& 0.8831
	\\
\end{tabular}
\end{center}
\end{table}

Classification ANNs with acceptable accuracy with respect to the validation dataset are obtained for the 16-, 24-, and even for the 32-dimensional ROM. These ANNs, denoted as $\tilde{\chi}^\mathrm{R16/24/32}$ in Tab.~\ref{tab_ann_class}, offer, in principle, the opportunity for an adaptive ROM scheme, in which for a given effective strain the lowest-dimensional but still acceptable ROM can be automatically identified for the chosen tolerances. In addition to the error classification of different ROMs, an attempt to classify the quality of the ANN labeled $\busig^{\mathrm{ANN1}}$ in Tab.~\ref{tab_sigma_anns} is made for the same tolerances with $\tilde\chi^\mathrm{ANN1}$, see last row in Tab.~\ref{tab_ann_class}. The surrogate $\busig^{\mathrm{ANN1}}$ has already intrinsic information of the training dataset, due to its optimization in respect to this dataset. In order to avoid an over-calibration, the training, validation and Monte Carlo datasets have been concatenated, randomly reordered and split into new training and validation datasets containing 90\% and 10\% of the data, respectively. The classifier $\tilde\chi^\mathrm{ANN1}$ for $\busig^{\mathrm{ANN1}}$ is trained on these new datasests. An extensive architecture test is performed with the same parameters as for the ROMs. The classifier for $\busig^{\mathrm{ANN1}}$, denoted as $\tilde{\chi}^\mathrm{ANN1}$ in Tab.~\ref{tab_ann_class}, reaches acceptable accuracy, but notably lower than the ones achieved for the ROM classifiers. In retrospective, a justification for the lower performance of the classifier $\tilde\chi^\mathrm{ANN1}$ is found in the higher regularity of the ROM solution that is matching the behavior of the full order model. This is explained by the ROM inheriting the mathematical structure and the physical principles of the FOM. The classifiers of this section allow for an on-the-fly model switching, as illustrated in Fig.~\ref{fig_flowchart}, to be exemplified in the following section. Hereby, the ROM32 is considered for Algorithm~\ref{algo:fe2r-nn:a} and Algorithm~\ref{algo:fe2r-nn:b} due to its sufficient accuracy, see Fig.~\ref{fig_error_distributions}.

%FF
\subsection{Multiscale simulation based on adaptive ANN-ROM-scheme}
\subsubsection{Twoscale problem}
The presented hybrid methods introduced in Algorithms~\ref{algo:fe2r-nn:a} and \ref{algo:fe2r-nn:b} are used in actual three-dimensional twoscale simulations. The results are compared to FE\textsuperscript{2R} simulations \citep[in the spirit of][]{Fritzen2016} in which the reduced order model is used as a stress surrogate in all points of a macroscopic structure which is considered as a reference based on the high accuracy of the ROM with 32 modes (see Fig.~\ref{fig_error_distributions}, Section~\ref{sec:ann:classifier}).

The macroscopic problem $(\bar{\mathrm{P}})$ depicted in Fig.~\ref{fig_fe2} is borrowed from \cite{Fritzen2016}, while the microstructure is described through the material of Sect.~\ref{sec_micro_mat}. Three different 3-dimensional mesh densities are considered on the macroscopic level: M1 (1,734 elements/13,872 int. points), M2 (6,318 elements/50,512 int. points) and M3 (53,790 elements/430,320 int. points). All three models consist of trilinear hexahedral elements with selectively reduced integration (i.e., B-bar elements are used). The loading in terms of a 2\% stretch of the macroscopic specimen is applied in 10 equally spaced increments up to 0.2\% followed by nine increments of 0.2\% amplitude each in order to better cover the transition between elastic and elasto-plastic  behavior for Algorithm~\ref{algo:fe2r-nn:b}. 

%%%%%%%%%%%%%%%%%%%%%%%%%%%%%%%%%%%%%%%%%%%%%%%%%%%%%%%%%%%%%%%%%%%%%%%%%%%
\subsubsection{Staggered adaptive procedure cf. Algorithm~\ref{algo:fe2r-nn:a}}
\label{sec:fe2r:algo:a}
%%%%%%%%%%%%%%%%%%%%%%%%%%%%%%%%%%%%%%%%%%%%%%%%%%%%%%%%%%%%%%%%%%%%%%%%%%%
First the staggered procedure introduced in Algorithm~\ref{algo:fe2r-nn:a} is used. It is found that the first run that is relying on the ANN surrogate only achieves excellent runtimes when evaluating the ANN on graphics cards (here: one Nvidia GTX Titan Black), leading to runtimes of approximately 15 seconds for one evaluation of the surrogate at each of the 430,320 integration points of the finest mesh M3. It shall be noted that this includes a major execution overhead\footnote{For simplicity each evaluation launches a new Python instance, reloads the model from a file and returns the results to the FE code through another file.}.

\begin{figure}[h]
 \centering
 \includegraphics[scale=1]{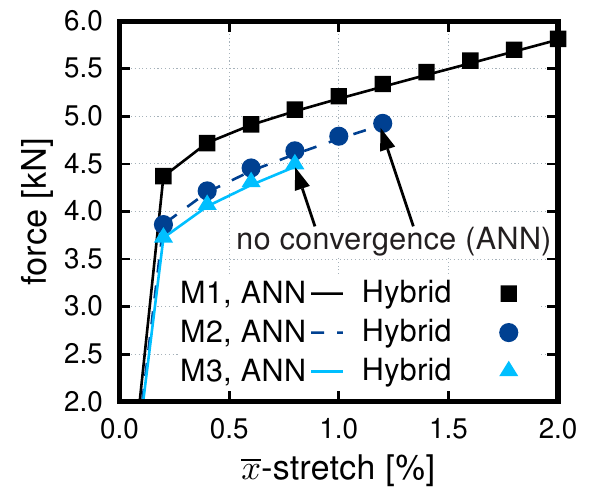}\quad
 \includegraphics[scale=1]{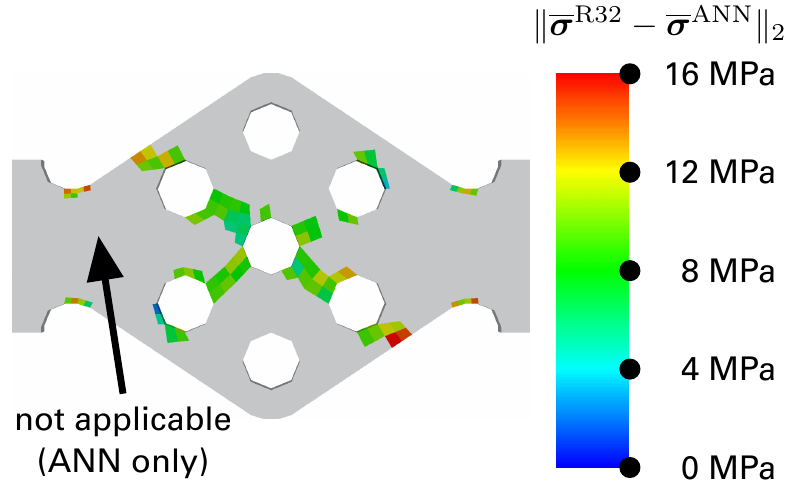}
 \caption{{\it left:} Comparison of the force of the ANN-only run (lines) and of the subsequent hybrid ANN/ROM run (symbols) for M1 (1734 elements), M2 (6318 elements) and M3 (53790 elements); {\it right:} Absolute error in the stress at the end of the hybrid run $\Vert \fbsigma^{\rm R32} - \fbsigma^{\rm ANN} \Vert_2$ for M1}
 \label{fig:force:comp:algo1}
\end{figure}

A general dilemma of twoscale simulations that was observed for the FE\textsuperscript{2R} method by \citet{Fritzen2016} is also present here: Local outliers of the strain field attain magnitudes that quickly exceed  the range of the inputs used during training of the ANN stress surrogate. The number of outliers becomes more relevant for the finer discretizations. This reveals a major short-coming of Algorithm~\ref{algo:fe2r-nn:a}: While the simulation for mesh level M1 terminated cleanly in roughly 3h wall-clock time with most of the computing time being spent during the hybrid ANN/ROM phase, M2 did not converge for loadings larger than 1.2\% due to locally excessive strains that lead to spurious stress response of the ANN. The finer mesh M3 fails to converge beyond 0.8\% of overall stretch. Additionally, the ANN version failed to improve the accuracy beyond a certain limit, i.e. it failed to achieve quadratic convergence starting beyond a critical load amplitude. In Fig.~\ref{fig:force:comp:algo1} a comparison of the tension force of the ANN-only run (lines) and of the subsequent hybrid run (symbols) is shown. During the hybrid run the number of integration points evaluating the ROM are determined from the quality indicator at the end of the last load step of the first run. For M1 this amounts to 960 out of 13,872 integration points (6.92\%)\footnote{The numbers for M2 and M3 are not representative as the final load was not achieved.}. These numbers illustrate that the ROM must be evaluated approximately 42,000 times for M1 (44 Newton iterations were needed in total) which leads to a substantial computational effort. Surprisingly, the ANN-only run and the hybrid run are hard to distinguish from the overall force-stretch plots, i.e. the ANN appears to yield good accuracy for this test. This is supported by the rather small absolute errors in the effective stress tensor, see Fig.~\ref{fig:force:comp:algo1} (right) for the final load and mesh M1. In summary, the staggered procedure can exclusively be used if the peak strain in the macroscopic problem is sufficiently low due to the aforementioned convergence issues. Then the solution can be expected to give accurate predictions.

In view of the number of quadrature points marked for use of the more reliable ROM, the adaptive scheme shows a steady increase when using the kinematic indicator~$\chi^{\rm K}$ marking points outside of the training range as not trustworthy for the ANN.

%%%%%%%%%%%%%%%%%%%%%%%%%%%%%%%%%%%%%%%%%%%%%%%%%%%%%%%%%%%%%%%%%%%%%%%%%%%
\subsubsection{Single pass on-the-fly adaptive algorithm cf. Algorithm~\ref{algo:fe2r-nn:b}}
\label{sec:fe2r:algo:b}
\begin{figure}[h]
 \centering
 \includegraphics[scale=1]{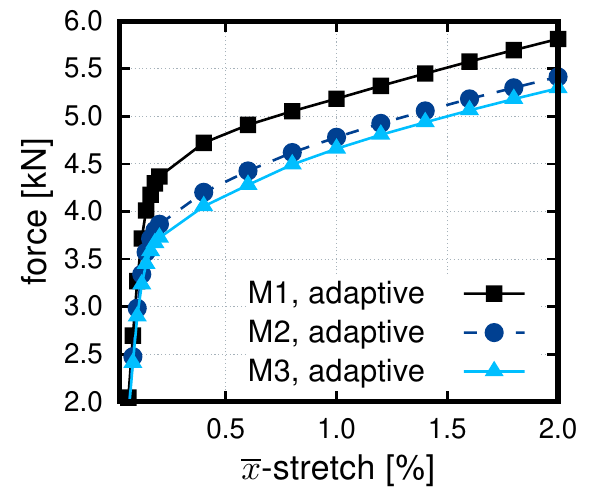}\quad
 \includegraphics[scale=1]{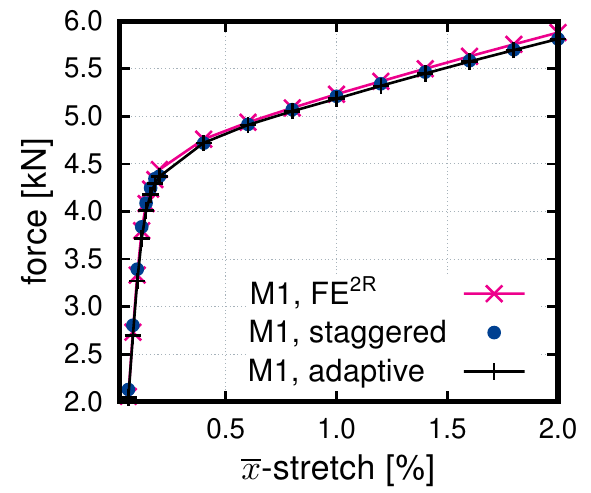}
 \caption{{\it left:} Comparison of the macroscopic stretch-force relation for the on-the-fly adaptive algorithm for meshes M1--M3; {\it right:} Comparison for M1: on-the-fly adaptive vs. staggered vs. FE\textsuperscript{2R} with 32 modes}
 \label{fig:force:comp:algo2}
\end{figure}

The crucial ingredient of the on-the-fly adaptive scheme, described in Algorithm~\ref{algo:fe2r-nn:b}, is the irreversible update of the quality indicator during each load increment. Thereby, alternating model selection is prevented. All three macroscopic models, M1, M2 and M3, converged without any issues. The resulting macroscopic tension force of all three is compared in Fig.~\ref{fig:force:comp:algo2}, where the hybrid curve from Fig.~\ref{fig:force:comp:algo1} and the FE\textsuperscript{2R} curve for the ROM featuring 32 modes are also shown for M1. It is observed from Fig.~\ref{fig:force:comp:algo2} (right) that all algorithms yield virtually identical results. Closer inspection reveals, however, that the FE\textsuperscript{2R} and adaptive algorithm have nearly indistinguishable slopes (despite a negligible shift), whereas the ANN model is slightly curved, i.e. it shows a qualitative difference towards the reference solution which gets more pronounced at increasing load amplitude.

\begin{figure}[h]
 \centering
 \includegraphics[scale=1]{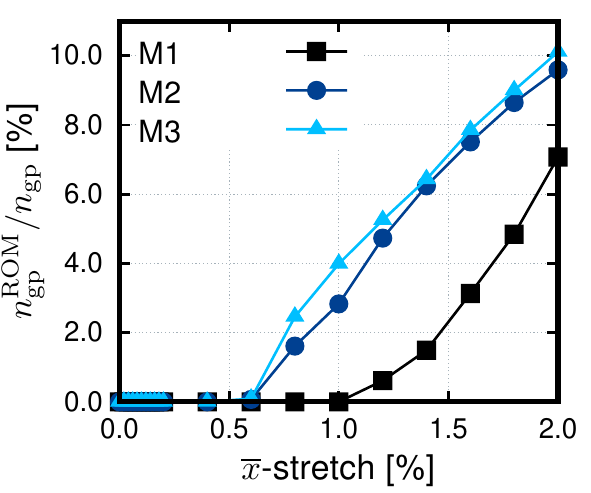}\quad
 \includegraphics[scale=1]{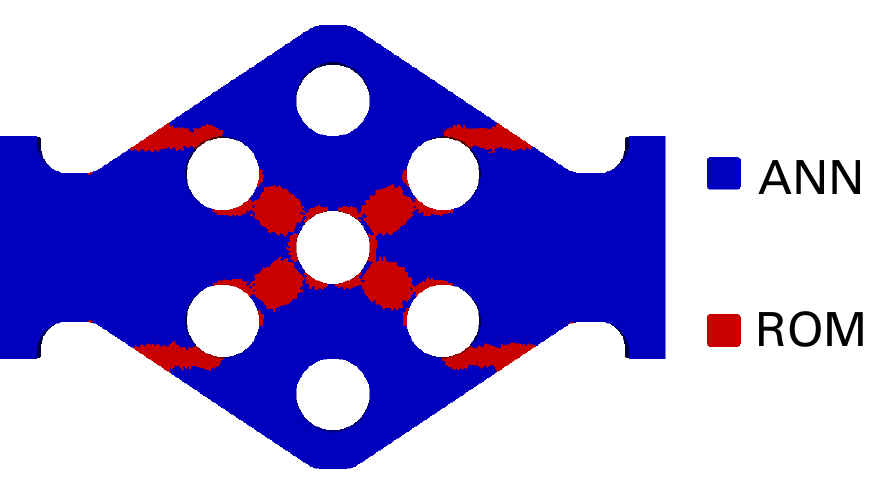}
\caption{{\it left:} Relative amount of integration points with active ROM during the adaptive twoscale simulation;
{\it right:} Quality indicator~$\chi^{\rm K}$ at the end of the adaptive twoscale simulation using mesh M3
}
 \label{fig:fe2r-nn:algo2:ngp}
\end{figure}

The adaptive algorithm has the advantage that the number of macroscopic integration points that require evaluation of the ROM depends only on the current state. For the considered proportional loading, and when using the kinematic indicator~$\chi^{\rm K}$, the relative amount of integration points grows monotonically with increasing load, cf. Fig.~\ref{fig:fe2r-nn:algo2:ngp}.

\begin{figure}[!h]
 \centering
 \includegraphics[scale=1]{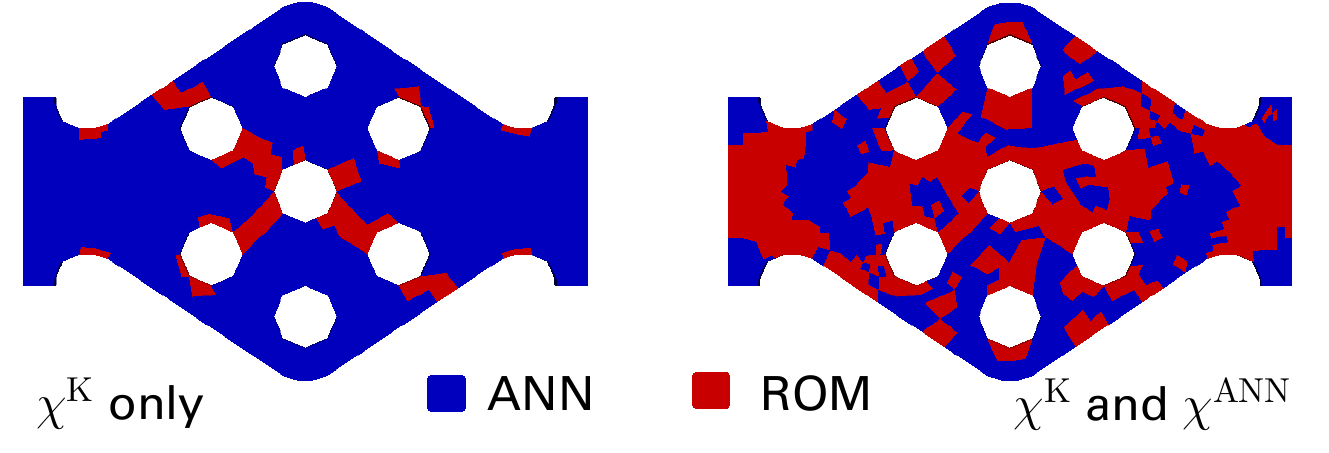}
 \caption{Comparison of the final quality indicators at the end of the adaptive simulation for mesh M1: kinematic classifier only {\it (left)} vs. hybrid classifier accounting for the ANN accuracy cf. Section~\ref{sec:ann:classifier} {\it (right)}}
 \label{fig:quality:M1}
\end{figure}

In order to investigate the practical usefulness of the ANN classifier, a comparison of the on-the-fly adaptive simulation using the kinematic quality indicator and the same simulation supplemented by the ANN classifier discussed in Section~\ref{sec:ann:classifier} is considered. As expected, the solid yet not overly satisfying accuracy of the ANN classifier (see Tab.~\ref{tab_ann_class}) induces a large number of additional ROM evaluations, thereby increasing the computation time considerably (approx. by a factor of~7), see Fig.~\ref{fig:quality:M1}. Notably, the ANN classifier adds a considerable amount of points at the left and right constriction and close to the holes. In order to assess the relevance of these additional points, the macroscopic tensile force was investigated: It varies less than 0.3\%, except in the very first load step with a difference around 0.6\%.

%%%%%%%%%%%%%%%%%%%%%%%%%%%%%%%%%%%%%%%%%%%%%%%%%%%%%%%%%%%%%%%%%%%%%%%%%%%

%FF+MF
\section{Concluding summary}
\label{sec_disscussion}

A multi-fidelity approach for generating surrogate models of the effective stress tensor for the use in twoscale simulations is developed in Section~\ref{sec:Twoscale}. At first, a ROM is derived from data gathered during full field simulations. The estimation of the error in the effective stress tensor (representing the QoI) of the ROM is discussed from a theoretical perspective in Section~\ref{sec:error:theory}. The mathematical structure of the error estimate reveals, that the ROM error estimation produces computational cost that is almost equivalent or even beyond that needed to solve a more dedicated ROM, thereby making it hard to justify such estimates when in the need for computational efficiency. 

In our view this dilemma can only be resolved by finding alternative surrogates with low computational complexity but moderate to good accuracy complemented by adaptive strategies for local model refinement that employ costly computational methods only when needed. In this regard, ANNs are seen as promising candidates for the calibration of surrogate models for the effective stress and for classification that can trigger adaptive refinement. In Section~\ref{sec:ANN}, the layout and the theoretical background of ANNs are discussed, together with different feature designs for the inputs and outputs based on the mechanical nature of the strain and stress. For the calibration of the stress surrogate, the mean squared error is used as loss function, while the quality of the trained ANN is checked on the validation dataset with the mean coefficient of determination and the mean relative norm error. In the case of error regression, a penalized mean squared error is proposed, which allows the conservative calibration of trained ANNs. For the error classification based on prescribed tolerances, the weighted cross entropy is used in order to allow for a better focus on the more important warning case, if the warning case density is low. Based on the proposed models, the core contribution of the present work constitutes two model-adaptive algorithms which encompass convergence issues encountered in the naive implementation of on-the-fly adaptive surrogate selection, see Section~\ref{sec:Adaptive}. The first staggered algorithm is based on a two run approach, in which the first run is conducted solely with the ANN effective stress surrogate and flags points evaluated outside of the strain training region, such that only these points are evaluated with the high-accuracy ROM in a second run. The second algorithm offers a more flexible on-the-fly model-adaptive approach by allowing the re-initialization of the ANN at the beginning of each load increment.

Numerical examples of the illustrated approaches are presented in Section~\ref{sec_nex} for a three-phase pseudo-plastic material with microstructure. First, ANNs are trained in order to approximate the effective stress. The surrogate of choice, $\busig^\mathrm{ANN1}$, achieved a mean relative norm error of 0.0189 and a mean coefficient of determination of 0.9995 and yields an accurate tangent stiffness, due to its formulation on the automatic differentiation capabilities of the TensorFlow library. The accuracy of the ANN stress surrogate is found to range between ROMs of dimension 24 and 32, respectively (see Section~\ref{sec:ANNvsROM}). The ansatz for error regression of ROMs of different dimension is presented, showing the possibilities for a calibration of a conservative ROM error estimator. In view of subsequent twoscale simulations with adaptive model selection, error classification is carried out for ROMs of different dimension and for the trained ANN stress surrogate. The achieved accuracy of the ROMs are higher than for the ANN stress surrogate, which indicates that the physics-informed ROM still shows a clearer pattern than the trained ANN stress surrogate, due to its inherited mathematical structure and underlying physical principles.

The trained ANNs are then used in twoscale mechanical FE simulations, based on the two developed algorithms of Section~\ref{sec:Adaptive}. The staggered algorithm produces sensible results but has two limitations: First, the number of macroscopic quadrature points marked for correction grows irreversibly. Second, the ANN surrogate must be sufficiently robust and of---at least---moderate accuracy in a prohibitive part of the strain space. This requirement stems from the fact that local strain outliers lead to queries that are way outside of the usual training range of the ANN. This effect is found to be more pronounced when the macroscopic mesh density is increased which further complicates the robust surrogate construction using purely data-driven methods in general, see~Section~\ref{sec:fe2r:algo:a}. The second algorithm offers a true on-the-fly adaptivity in which the ANN surrogate can be recovered, e.g., during unloading. It is observed in Section~\ref{sec:fe2r:algo:b} that this second algorithm offers the fastest convergence among the considered twoscale simulations being approximately 3-10 times faster than the staggered algorithm and around 20~times in comparison to the fully coupled FE\textsuperscript{2R} algorithm using the ROM with 32 modes for all stress predictions. The adaptive on-the-fly model of the second algorithm offers, therefore, an attractive approach which combines a low number of ROM evaluations with good convergence. 

The final test using the additional error classifier for the ANN stress surrogate introduced a high number of additional negative outcomes (i.e., ANN error greater than tolerances), considerably increasing the number of integration points requiring the ROM. This was expected due to the low accuracy achieved during the training of the classifier, more specifically, due to the low accuracy for the positive outcome ACC$_1$ and corresponding high amount of positive outcomes reflected by $w_0$, see Tab.~\ref{tab_ann_class}. On the one hand, this last approach offers a conservative twoscale scheme relying on the robust ROM. On the other hand, the approach loses computational efficiency, which leaves the error classification for rapid multiscale problems still an open issue. Future improvements should, therefore, focus on further theoretical or hybrid error estimators and an improved error classification for the effective stress. The latter could benefit from datasets spanning larger portions of the strain space. The authors are convinced that the ambitious goal of reliable twoscale simulations can only be achieved by fusing data-driven methods together with dedicated theories. For example, the reader should consider that at least two quality indicators are indeed necessary for a reliable model-adaptive ansatz. One quality indicator should address the trained ANN stress surrogate over the sampled strain (input) space and one quality indicator should definitely return a warning if the input leaves the training region. This is quintessential, since the range of macroscopic strain is not a priori known and the behavior of the purely data-driven ANN outside the training region may endanger the convergence and quality of the multiscale simulation. The results of the present work support the usability of ANNs in computational materials science, although further research ideally addressing the combination of theory and data is urgently required. 

%%%%%%%%%%%%%%%%%%%%%%%%%%%%%%%%%%%%%%%%%%%%%%%%%%%%%%%%%%%%%%%%%%%%%%%%%%%

\section*{Conflict of Interest Statement}

The authors declare that the research was conducted in the absence of any commercial or financial relationships that could be construed as a potential conflict of interest.

\section*{Author Contributions}

FF conducted the finite element and reduced order model computations on the microscale and the twoscale simulations based on his self-developed code and wrote the corresponding sections. MF implemented, trained the artificial neural networks and wrote the corresponding sections. FL conducted together with FF preliminary work on theoretical error estimates solely based on the reduced order model, which yielded the theoretical insights for the computational expenses and the necessity for alternative approaches. FL wrote the corresponding section for the theoretical error estimation. 

\section*{Funding}
\addcontentsline{toc}{section}{Funding}
The contributions of MF and FF are funded by the Deutsche Forschungsgemeinschaft (DFG, German Research Foundation) -- FR2702/6 – within the scope of the Emmy Noether Group EMMA -- Efficient Methods for Mechanical Analysis.
The contributions of Fredrik Larsson are funded by the Swedish Research Council (VR) under grant no. 2015-05422.

\section*{Acknowledgments}
\addcontentsline{toc}{section}{Acknowledgments}
Vivid discussions within the scope of Cluster of Excellence SimTech (DFG EXC310 and EXC2075) regarding machine learning and data-driven model surrogation are highly appreciated. Felix Fritzen and Mauricio Fern\'{a}ndez further acknowledge the valuable discussions with Steffen Freitag (Ruhr-Universit\"at-Bochum) on the topic of ANN-based regression and classification.

\section*{\setword{Supplementary Material}{supp_mat}}
\addcontentsline{toc}{section}{\ref{supp_mat}}
The Supplementary Material for this article can be found online at: \url{https://www.frontiersin.org/articles/10.3389/fmats.2019.00075/full#supplementary-material}.

\textbf{\setword{Supplemental Data (Data sheet 1.zip)}{supp_data}}: Supplementary material is provided in the form of two HDF5 datasets (containing all FEM and ROM results used for the ANN training). Further, the stress surrogate $\bar\usig^\mathrm{ANN1}$ and the quality indicator for this ANN are provided including a Python interface for accessing the file.

\end{document}